\documentclass[review]{elsarticle}
\usepackage{geometry}
\usepackage{amsmath}
\usepackage{amsfonts}
\usepackage{amssymb}
\usepackage{graphicx}
\usepackage{dcolumn}
\usepackage{bm}
\usepackage[colorlinks = true, allcolors = blue]{hyperref}
\usepackage{float}
\usepackage{subcaption}

\makeatletter
\def\ps@pprintTitle{%
 \let\@oddhead\@empty
 \let\@evenhead\@empty
 \def\@oddfoot{}%
 \let\@evenfoot\@oddfoot}
\makeatother
\begin{document}
\newcommand{\bd}{\begin{document}}
\newcommand{\ed}{\end{document}}
\newcommand{\bc}{\begin{center}}
\newcommand{\ec}{\end{center}}
\newcommand{\bfr}{\begin{flushright}}
\newcommand{\efr}{\end{flushright}}
\newcommand{\lt}{\left}
\newcommand{\rt}{\right}
\newcommand{\vs}{\vspace}
\newcommand{\hs}{\hspace}
\newcommand{\beq}{\begin{equation}}
\newcommand{\eeq}{\end{equation}}
\newcommand{\lb}{\linebreak}
\newcommand{\pb}{\pagebreak}
\newcommand{\mb}{\makebox}
\newcommand{\fb}{\framebox}
\newcommand{\mc}{\lamlticolumn}
\newcommand{\ben}{\begin{enumerate}}
\newcommand{\een}{\end{enumerate}}
\newcommand{\bit}{\begin{itemize}}
\newcommand{\eit}{\end{itemize}}
\newcommand{\oln}{\overline}
\newcommand{\un}{\underline}
\newcommand{\lefq}{\lefteqn}
\newcommand{\ba}{\begin{array}}
\newcommand{\ea}{\end{array}}
\newcommand{\beqa}{\begin{eqnarray}}
\newcommand{\eeqa}{\end{eqnarray}}
\newcommand{\beqas}{\begin{eqnarray*}}
\newcommand{\eeqas}{\end{eqnarray*}}
\newcommand{\bfg}{\begin{figure}}
\newcommand{\efg}{\end{figure}}
\newcommand{\bds}{\begin{displaymath}}
\newcommand{\eds}{\end{displaymath}}
\newcommand{\btb}{\begin{tabbing}}
\newcommand{\etb}{\end{tabbing}}
\newcommand{\para}{\parallel}
\newcommand{\pad}{\partial}
\newcommand{\nn}{\nonumber}
\newcommand{\la}{\leftarrow}
\newcommand{\ra}{\rightarrow}
\newcommand{\lgla}{\longleftarrow}
\newcommand{\lgra}{\longrightarrow}
\newcommand{\La}{\Leftarrow}\newcommand{\Ra}{\Rightarrow}
\newcommand{\Lra}{\Leftrightarrow}
\newcommand{\Lgla}{\Longleftarrow}
\newcommand{\Lgra}{\Longrightarrow}
\newcommand{\lan}{\langle}
\newcommand{\ran}{\rangle}
\renewcommand{\a}{\alpha}
\renewcommand{\b}{\beta}
\newcommand{\g}{\gamma}
\newcommand{\G}{\Gamma}
\renewcommand{\d}{\delta}
\newcommand{\eps}{\epsilon}
\newcommand{\Th}{\Theta}
\newcommand{\s}{\sigma}
\newcommand{\lam}{\lambda}
\newcommand{\D}{\Delta}
\newcommand{\ds}{\displaystyle}
\newcommand{\vare}{E}
\newcommand{\pr}{\prime}
\newcommand{\ro}{\rho}
\newcommand{\nab}{\nabla}
\newcommand{\m}{\lam}
\newcommand{\n}{\nu}
\newcommand{\Sg}{\Sigma}
\newcommand{\p}{\pi}
\newcommand{\R}{I\!\!R}
\newcommand{\om}{\omega}
\newcommand{\Om}{\Omega}
\newcommand{\ovra}{\overrightarrow}
\newcommand{\ze}{\zeta}
\newcommand{\vart}{\vartheta}
\newcommand{\tri}{\triangle}
\newcommand{\f}{\frac}
\newcommand{\iny}{\infty}
\newcommand{\pro}{\propto}
\renewcommand{\arraystretch}{1.25}

\begin{frontmatter}
	\title{Time dependent rationally extended P\"oschl-Teller potential and some of its properties}
	\author[ad1]{D.~Nath}
	\ead{debrajn@gmail.com} 

	\author[ad2,ad3]{P.~Roy}
	\ead{Corresponding author: pinaki.roy@tdtu.edu.vn}
	\address[ad1]{Department of Mathematics, Vivekananda College, 269 D.H. Road, Kolkata - 700063, India.}
	\address[ad2]{Atomic~Molecular~and~Optical~Physics~Research~Group, Advanced Institute of Materials Science, Ton Duc Thang University, Ho~Chi~Minh~City, Vietnam}
	\address[ad3]{Faculty of Applied Sciences, Ton Duc Thang University, Ho~Chi~Minh~City,~Vietnam}

\begin{abstract}
We examine time dependent Schr\"odinger equation with oscillating boundary condition. More specifically, we use separation of variable technique to construct time dependent  rationally extended P\"oschl-Teller potential (whose solutions are given by in terms of $X_1$ Jacobi exceptional orthogonal polynomials) and its supersymmetric partner, namely the P\"oschl-Teller potential. We have obtained exact solutions of the Schr\"odinger equation with the above mentioned potentials subjected to some  boundary conditions of the oscillating type. A number of physical quantities like the average energy, probability density, expectation values etc. have also been computed for both the systems and compared with each other.
\end{abstract}
\begin{keyword}
Exceptional orthogonal polynomials; oscillating boundary condition; exact solutions
\end{keyword}
\end{frontmatter}

\section{Introduction}
Schr\"odinger equation with time dependent potentials are of interest for various reasons and they have been studied over the years by many authors \cite{TD.Schrodinger,TD.Schrodinger.2,TD.Schrodinger.3,TD.Schrodinger.4,TD.Schrodinger.5,TD.Schrodinger.6,TD.Schrodinger.7,TD.Schrodinger.8,TD.Schrodinger.9}. Among the various time dependent systems, there is a class of problems for which the boundary moves with time \cite{Fermi,MB.First,moving.wall,moving.wall.2,moving.wall.3,moving.wall.4,moving.wall.5,moving.wall.6,moving.wall.7,moving.wall.8,moving.wall.9,moving.wall.10,moving.wall.11,moving.wall.12,moving.wall.13,moving.wall.14,moving.wall.15,moving.wall.16,moving.wall.17,moving.wall.18}. During the last few years moving boundary problems have been studied by various methods like the method of invariants \cite{invariant}, separation of variables \cite{symmetry.separation,symmetry.separation.2,symmetry.separation.3,symmetry.separation.4}, supersymmetry \cite{TD.Schrodinger.6,moving.wall.14} etc. It may be noted that in most of these works the initial potential at zero time was chosen as one of the standard solvable quantum mechanical potentials.
 
The solutions of the standard quantum mechanical potentials are usually given in terms of classical orthogonal polynomials. On the other hand, some years back a class of orthogonal polynomials, called the exceptional orthogonal polynomials have been found and they are distinct from the classical ones \cite{eop,eop1}. Subsequently through Darboux transformation or supersymmetry a large number of exactly solvable potentials have been found whose solutions are given in terms of exceptional orthogonal polynomials \cite{eseop,dutta}. One of the main features of these new potentials is that they share the same or nearly the same spectrum as their classical counterparts although their wavefunctions are quite different. Recently, a method has also been proposed to generate rational extensions of time dependent potentials \cite{axel}. In the present work we shall consider the P\"oschl-Teller potential and its supersymmetric partner, namely the rationally extended P\"oschl-Teller potential \cite{quesne} and apply the separation of variable technique \cite{symmetry.separation.4} to construct time dependent versions of both the  potentials subjected to oscillating boundary condition \cite{nieto}.  One of the reasons for studying this problem stems from the fact that barring an exception \cite{axel} time dependent versions of  potentials whose solutions are given in terms of exceptional orthogonal polynomials have been not been studied before and secondly, it is of interest to examine how far physical quantities like average energy or expectation values etc. for the time dependent rationally extended P\"oschl-Teller potential differ from their time dependent classical counterpart, namely the time dependent P\"oschl-Teller potential. In particular, we shall obtain exact solutions and examine various features like behavior of the probability density, time dependent localization property, average energy, root mean square values etc. and also compare these properties with those of the time dependent P\"oschl-Teller potential.

The paper is organized as follows: in Section 2 we present briefly the separation of variable approach to the time dependent Schr\"odinger equation; in Section 3 we shall use the method of section 2 and supersymmetry to construct time dependent rationally extended P\"oschl-Teller potential. We shall also  examine various features of the time dependent P\"oschl-Teller potential and the rationally extended P\"oschl-Teller potential and compare the differences, if any, at different points of time; finally, Section 4 is devoted to a conclusion.
\section{Separation of variables}
To begin with, let us consider the time dependent Schr\"odinger equation (in units of $\hbar=2m=1$)
\beq
\ds\left[-\displaystyle\f{\pad^{2}}{\pad x^{2}}+V(x,t)\right]\psi(x,t)=\ds i\displaystyle\f{\pad\psi(x,t)}{\pad t},\label{sxt}
\eeq
subjected to the boundary conditions
\beq
\ds\psi(0,t)=0=\psi\left(L(t),t\right).
\eeq
The above boundary condition describes a moving boundary \cite{moving.wall} problem in quantum mechanics. Depending on the choice of $L(t)$ the boundary can be either expanding, contracting or oscillating.
We shall now use the separation of variable technique \cite{symmetry.separation} to transform Eq.(\ref{sxt}) to a problem with fixed boundary. To do this let us transform the variable $x\mapsto q$ as
\beq\label{xq}
q=\ds\f{\pi\left(x-\a(t)\right)}{L(t)},~\a(t)=\f{1}{2}L(t),~-\ds\f{\pi}{2}\leq q\leq \f{\pi}{2}.
\eeq
We now consider the potential and the wavefunction to be of the form
\beq
\ds V\big(x(q,t),t\big)=\ds g(t)\widetilde{V}(q)+U(q,t)+\ds g_{0}(t),\label{vxt}
\eeq 
\beq\label{wf}
\ds\psi(q,t)=e^{\Phi(q,t)}~Q(q)T(t).
\eeq

Then substituting (\ref{wf}) into Eq. (\ref{sxt}) we obtain
\beq
-\ds\f{1}{Q}\f{d^{2}Q}{dq^{2}}+L_{1}^{2}(t)\left[g(t)\widetilde{V}(q)\right]=iL_{1}^{2}(t)\f{1}{T}\f{dT}{dt}=\eps,\label{sepa}
\eeq
where $\eps$ is a separation constant and
%
\beq\label{q.abc}
\ba{l}
\ds\Phi(q,t)=a(t)\f{q^{2}}{2}+b(t)q+c(t),~~~~\ds a(t)=\f{i}{2}L_{1}(t)\dot{L}_{1}(t),\\
\ds b(t)=\f{i}{2}L_{1}(t)\dot{\a}(t),~~~~\ds L_{1}(t)=\f{1}{\pi}L(t),
\ea
\eeq
and $c(t)$ is a constant of integration. Now without any loss of generality we can choose \cite{symmetry.separation.4}
\beq
\ds c(t)=-\f{1}{2}\ln L_{1}(t)+\f{i}{4}\int_{0}^{t}\dot{\a}^{2}(s)ds-i\int_{0}^{t}g_{0}(s)ds,\label{ct}
\eeq

\beq\label{uxt}
\ds U(q,t)=-\f{1}{4}\big(\pi q+q^{2}\big)L_{1}(t)\ddot{L}_{1}(t),
\eeq
and 
\beq
\ds g(t)L_{1}^{2}(t)=\mbox{constant }=1, \mbox{say}.
\eeq
Then separating Eq. (\ref{sepa}) into the space dependent and the time dependent part we obtain
\beq\label{Eq.Qq}
-\ds\f{d^{2}Q}{dq^{2}}+\widetilde{V}(q)Q=\eps Q,
\eeq
and
\beq\label{Eq.Tt}
\ds T(t)=e^{-i\eps \tau (t)},~~~\tau (t)=\int_{0}^{t}\f{1}{L_{1}^{2}(s)}ds,
\eeq
where $\widetilde{V}(q)$ is the effective potential. Finally from Eqs. (\ref{vxt}) and (\ref{uxt}) the time dependent potential can be found to be
\beq
\ds V(x,t)=\f{\pi^{2}}{L^{2}(t)}\widetilde{V}\left[\f{\pi\left(x-\f{1}{2}L(t)\right)}{L(t)}\right]+\f{1}{16}L(t)\ddot{L}(t)-\f{1}{4}\f{\ddot{L}(t)}{L(t)}x^{2}+g_{0}(t).
\eeq

\subsection{Time dependent rationally extended P\"oschl-Teller potential via supersymmetry}
Here we shall construct a potential appropriate for the moving boundary problem using supersymmetry formalism \cite{susy1, susy2}. As mentioned earlier here we are interested in obtaining time dependent potentials whose time independent counterparts are solvable in terms of exceptional orthogonal polynomials. Thus we consider the following superpotential \cite{quesne}
\beq
\ds\widetilde{W}(q,A,B)=(-B-\f{1}{2})\tan q + (A-\f{1}{2})\sec q + \f{2B\cos q}{2A-1-2B\sin q}.
\eeq
Then the supersymmetric partner potentials ${\widetilde V}^{\pm}(q,A,B)={\widetilde W}^2\pm {\widetilde W}^\prime$ can be found to be
\beq
\ba{ll}
\ds \widetilde{V}^{(-)}(q,A,B)&=\left[A(A-1)+(B+1)^{2}\right]\sec^{2}q-(B+1)(2A-1)\sec q\tan q-\left(B-\f{1}{2}\right)^{2},\\
\ds\widetilde{V}^{(+)}(q,A,B)&=\left[A(A-1)+B^{2}\right]\sec^{2}q-B(2A-1)\sec q\tan q +\f{2(2A-1)}{2A-1-2B\sin q}\\
&~~~~\ds+\f{2\left[(2A-1)^{2}-4B^{2}\right]}{\left(2A-1-2B\sin q\right)^{2}}-\left(B-\f{1}{2}\right)^{2}.
\ea
\eeq
The eigenvalues and the corresponding wavefunctions are given by \cite{quesne}
\beq
\ds E_{n}^{(-)}=\big(n+A\big)^{2}-\big(B-\f{1}{2}\big)^{2}=E_{n}^{(+)},
\eeq 
\beq
\displaystyle Q_{n}^{(-)}(q)=N_{n}^{(\a-1,\b+1)}(1-z)^{\f{1}{2}(\a-\f{1}{2})}(1+z)^{\f{1}{2}(\b+\f{3}{2})}P_{n}^{(\a-1,\b+1)}(z),
\eeq
\beq
\ds Q_{n}^{(+)}(q)=\f{N_{n}^{(\a-1,\b+1)}}{\sqrt{E_{n}^{(-)}}}\f{(1-z)^{\f{1}{2}(\a+\f{1}{2})}(1+z)^{\f{1}{2}(\b+\f{1}{2})}}{\b+\a-(\b-\a)z}\widehat{O}^{(\a,\b)}P_n^{(\a-1,\b+1)}(z),
\eeq
where
\beq
\ba{l}
\widehat{O}^{(\a,\b)}P_n^{(\a-1,\b+1)}(z)=4(\b-\a)(\b+n)\widehat{P}_{n+1}^{(\a,\b)}(z)\\
\widehat{P}_{n+1}^{(\a,\b)}(z)=\ds\f{1}{2}\left(\f{\b+\a}{\b-\a}-z\right)P_{n}^{(\a,\b)}(z)+(\b+\a+2n)^{-1}\left(\f{\b+\a}{\b-\a}P_{n}^{(\a,\b)}(z)-P_{n-1}^{(\a,\b)}(z)\right),
\ea
\eeq
$\a=A-B-\f{1}{2}$, $\b=A+B-\f{1}{2}$, $z=\sin q$ and
\beq
\displaystyle N_{n}^{(\a,\b)}=\sqrt{\f{n!(2n+\a+\b+1)\Gamma(n+\a+\b+1)}{2^{\a+\b+1}\Gamma(n+\a+1)\Gamma(n+\b+1)}}\,,
\eeq
is the normalization constant and $P_{n}^{(\a,\b)}(z)$ denotes the Jacobi polynomial of degree $n$ in $z$ with parameters $(\a,\b)$. It may be noted that $\widehat{P}_{n+1}^{(\a,\b)}(z)$ which is a combination of Jacobi polynomials is called the $X_1$ Jacobi exceptional orthogonal polynomial \cite{quesne}. It may be noted that the $\widetilde{V}^{(-)}$ is a standard P\"oschl-Teller potential while its supersymmetric partner is solvable in terms of the exceptional orthogonal polynomials $\widehat{P}_{n+1}^{(\a,\b)}(z)$. Next in order to obtain the time dependent potentials we use Eq. (\ref{vxt}) to find
\beq
\ba{ll}\label{Pot.Vmxt}
\ds V^{(-)}(x,t)=&\f{\pi^{2}}{L^{2}(t)}\left[A(A-1)+(B+1)^{2}\right]\sec^{2}\left[\f{\pi\big(x-\f{1}{2}L(t)\big)}{L(t)}\right]\\
&\ds-~\f{\pi^{2}}{L^{2}(t)}(B+1)(2A-1)\sec\left[\f{\pi\left(x-\f{1}{2}L(t)\right)}{L(t)}\right]\tan\left[\f{\pi\left(x-\f{1}{2}L(t)\right)}{L(t)}\right]\\
&+~\f{1}{16}L(t)\ddot{L}(t)~-~\f{1}{4}\f{\ddot{L}(t)}{L(t)}x^{2}~-~\left(B-\f{1}{2}\right)^{2}.
\ea
\eeq
Now, using Eqs.(\ref{Eq.Qq}), (\ref{Eq.Tt}) we obtain the time dependent wavefunction as
\beq\label{psimxt}
\ds\psi^{(-)}_n(x,t)=\sqrt{\Om^{(-)}(x,t)}~\Phi^{(-)}_n(x,t)~e^{iF_n^{(-)}(x,t)},
\eeq
where
\beq
\ba{ll}
\ds\Om^{(-)}(x,t) 
&=\f{\pi}{L(t)}\cos^{2A} \left[\f{\pi\left(x-\f{1}{2}L(t)\right)}{L(t)}\right]\left(\f{1~+~\sin \left[\f{\pi\left(x-\f{1}{2}L(t)\right)}{L(t)}\right]}{1~-~\sin \left[\f{\pi\left(x-\f{1}{2}L(t)\right)}{L(t)}\right]}\right)^{B+1},
\ea
\eeq
\beq
\ds\Phi^{(-)}_n(x,t)=P_{n}^{(A-B-\f{3}{2},A+B+\f{1}{2})}\left(\sin \left[\f{\pi\left(x-\f{1}{2}L(t)\right)}{L(t)}\right]\right),
\eeq
and $e^{iF_n^{(-)}(x,t)}$ is the position and time dependent phase factor:
\beq
F_n^{(-)}(x,t)=\f{\dot{L}(t)}{4L(t)}x^{2}~-~\f{1}{16}L(t)\dot{L}(t)~+~\f{1}{16}\int_{0}^{t}\dot{L}^{2}(t)dt-E_{n}^{(-)}~\int_{0}^{t}\f{\pi^2}{L^{2}(s)}ds.
\eeq
It is not difficult to show that the normalization condition is satisfied :
\beq
\int_0^{L(t)}\Om^{(-)}(x,t)\Phi^{(-)}_n(x,t)\Phi^{(-)}_m(x,t)dx=\f{\d_{n,m}}{\left[N_n^{(A,B+1)}\right]^2}.
\eeq

Now we come to the time dependent rational extension of ${\tilde V^{(-)}}(q)$. From Eq. (\ref{vxt}) the time dependent potential $V^{(+)}(x,t)$ is found to be
\beq
\ba{ll}\label{Pot.Vpxt}
\ds V^{(+)}(x,t)=& \f{\pi^{2}}{L^{2}(t)}\left[A(A-1)+B^{2}\right]\sec^{2}\left[\f{\pi\left(x-\f{1}{2}L(t)\right)}{L(t)}\right]\\
&\ds -~\f{\pi^{2}}{L^{2}(t)}B(2A-1)\sec \left[\f{\pi\left(x-\f{1}{2}L(t)\right)}{L(t)}\right]\tan\left[\f{\pi\left(x-\f{1}{2}L(t)\right)}{L(t)}\right]\\
&\ds+\f{2(2A-1)}{2A-1-2B\sin\left[\f{\pi\left(x-\f{1}{2}L(t)\right)}{L(t)}\right]}-\f{2[(2A-1)^{2}-4B^{2}]}{\left(2A-1-2B\sin \left[\f{\pi\left(x-\f{1}{2}L(t)\right)}{L(t)}\right]\right)^{2}}\\
&\ds +~\f{1}{16}L(t)\ddot{L}(t)~-~\f{1}{4}\f{\ddot{L}(t)}{L(t)}x^{2}~-~\left(B-\f{1}{2}\right)^{2}.
\ea
\eeq
The corresponding wave function can be found using Eqs.(\ref{Eq.Qq}) and (\ref{Eq.Tt}): 
\beq\label{psipxt}
\ds\psi^{(+)}_n(x,t)= 
\sqrt{\Om^{(+)}(x,t)}~\Phi^{(+)}_n(x,t)~e^{iF_n^{(+)}(x,t)},
\eeq
where 
\beq
\ba{ll}
\ds\Om^{(+)}(x,t)
&=\f{1~-~\sin \left[\f{\pi\left(x-\f{1}{2}L(t)\right)}{L(t)}\right]}{1~+~\sin \left[\f{\pi\left(x-\f{1}{2}L(t)\right)}{L(t)}\right]}\times\f{\Om^{(-)}(x,t)}{\left(2A-1-2B\sin\left[\f{\pi\left(x-\f{1}{2}L(t)\right)}{L(t)}\right]\right)^2},
\ea
\eeq 
\beq
\ba{ll}
\ds \Phi^{(+)}_n(x,t)=&\f{1}{2}\left(\f{\b+\a}{\b-\a}-\sin \left[\f{\pi\left(x-\f{1}{2}L(t)\right)}{L(t)}\right]\right)P_{n}^{(\a,\b)}\left(\sin \left[\f{\pi\left(x-\f{1}{2}L(t)\right)}{L(t)}\right]\right)\\
\ds&+~(\b+\a+2n)^{-1}\f{\b+\a}{\b-\a}P_{n}^{(\a,\b)}\left(\sin \left[\f{\pi\left(x-\f{1}{2}L(t)\right)}{L(t)}\right]\right)\\
\ds&-~(\b+\a+2n)^{-1}P_{n-1}^{(\a,\b)}\left(\sin\left[\f{\pi\left(x-\f{1}{2}L(t)\right)}{L(t)}\right]\right),
\ea
\eeq 
are the time dependent exceptional orthogonal polynomials  and $e^{iF_n^{(+)}(x,t)}$ is a position and time dependent phase factor:

\beq
F_n^{(+)}(x,t)=\f{\dot{L}(t)}{4L(t)}x^{2}~-~\f{1}{16}L(t)\dot{L}(t)~+~\f{1}{16}\int_{0}^{t}\dot{L}^{2}(t)dt-E_{n}^{(+)}~\int_{0}^{t}\f{\pi^2}{L^{2}(s)}ds.
\eeq
The normalization condition reads 
\beq
\int_0^{L(t)}\Om^{(+)}(x,t)\Phi^{(+)}_n(x,t)\Phi^{(+)}_m(x,t)dx=\f{E_n^{(+)}}{4\left[(\b-\a)(\b+n)N_n^{(A,B+1)}\right]^2}\d_{n,m}
\eeq

\section{Expectation values}
In this section we shall evaluate several expectation values of interest for both the time dependent potentials (\ref{Pot.Vmxt}) and (\ref{Pot.Vpxt}). To this end we first note that the root mean square (RMS) or standard deviation is defined as
\beq
\left(\Delta x\right)=\sqrt{\langle x^2\rangle-\langle x\rangle^2},
\eeq
where
\beq
\langle x^k\rangle=\int \psi^*(x,t)\,x^k\,\psi(x,t)\,dx,~k=1,2,
\eeq 
is the $k$th order moment of $x$ and  $\psi^*(x,t)$ is the complex conjugate of $\psi(x,t)$. The RMS is the direct spreading measure of $x$ and it has unit of length $x$. It has a number of interesting properties of which we shall use the following ones: It is invariant under (i) translations and reflections, (ii) linear scaling \cite{RMS,RMS.2}. The solutions $\psi_n^{(\pm)}(x,t)$ of the time dependent potentials $V^{(\pm)}(x,t)$ are invariant neither under translations nor scaling, that is, under the transformation (\ref{xq}). On the other hand the probability densities $|\psi_n^{(\pm)}(x,t)|^2=\f{\pi}{L(t)}|Q_n^{(\pm)}(q)|^2$ are invariant under the above mentioned transformation. Moreover it is seen that the time dependent potential is shifted by two additional terms $U(q,t)$ and $g_0(t)$ which are also linearly scaled by the factor $\f{\pi^2}{L^2(t)}$. Then using the second property mentioned above we obtain
\beq\label{std.deviation}
\left(\Delta x\right)^2=\f{L^2(t)}{\pi^2}\left(I_2-I_1^2\right),
\eeq 
where
\beq
I_k=\int Q(q)^2\,q^k\,dq,~ k=1,2.
\eeq
From (\ref{std.deviation}) it may be seen that the RMS for any state of the time dependent potential can be obtained by linearly scaling by $\f{L(t)}{\pi}$ the RMS of corresponding state of the time independent potential. Next, the RMS in the momentum space is defined by
\beq
(\Delta p)=\sqrt{\langle p^2\rangle-\langle p\rangle^2},
\eeq 
where
\beq
\langle p^k\rangle=\int \psi^*\left(-i\f{\partial}{\partial x}\right)^k\psi dx=(-i)^k\left(\f{\pi}{L(t)}\right)^{k-1}\int \psi^*\f{\partial^k\psi}{\partial q^k} dq.
\eeq
After some calculations we finally obtain 
\beq\label{std.deviation.p}
(\Delta p)^2=\f{\pi^2}{L^2(t)}\left[I_3-a^2(t)\left(I_2-I_1^2\right)\right],
\eeq
where $I_3=\int \left(\f{\partial Q}{\partial q}\right)^2\, dq$. 
Therefore the Heisenberg uncertainty relation is
\beq
(\Delta x)(\Delta p)=\left\{\left(I_2-I_1^2\right)\left[I_3-a^2(t)\left(I_2-I_1^2\right)\right]\right\}^{\f{1}{2}}\ge \f{1}{2}.
\eeq
Now from Eqs. (\ref{std.deviation}), (\ref{std.deviation.p}) we may write
\beq
\left(\Delta x\right)_n^{(\pm)}=\f{L(t)}{\pi}\sqrt{I_{2,n}^{(\pm)}-\left[I_{1,n}^{(\pm)}\right]^2},
\eeq
\beq
(\Delta p)_n^{(\pm)}=\f{\pi}{L(t)}\sqrt{I_{3,n}^{(\pm)}-a^2(t)\left(I_{2,n}^{(\pm)}-\left[I_{1,n}^{\pm}\right]^2\right)}.
\eeq
Therefore the Heisenberg uncertainty relation can be expressed as 
\beq\label{uncer}
(\Delta x)_n^{(\pm)}(\Delta p)_n^{(\pm)}=\left\{\left(I_{2,n}^{(\pm)}-\left[I_{1,n}^{(\pm)}\right]^2\right)\left[I_{3,n}^{(\pm)}-a^2(t)\left(I_{2,n}^{(\pm)}-\left[I_{1,n}^{(\pm)}\right]^2\right)\right]\right\}^{\f{1}{2}}\ge \f{1}{2}.
\eeq

Next, we note that the average energy of a normalized time dependent state $\psi(x,t)$ is defined by \cite{average.energy}
\beq
\ba{ll}
\bar{E}=\langle H\rangle&=i\int \psi^*(x,t)\f{\partial \psi(x,t)}{\partial t}\,dx\\
&=i\ds\int Q(q)\left\{\left(i\dot{F}(t)-\f{\dot{L}_1(t)}{2L_1(t)}\right)Q(q)-\f{\dot{\a}(t)+q\dot{L}_1(t)}{L_1(t)}\left(\f{\partial Q}{\partial q}+iQ(q)\f{\partial F}{\partial q}\right)\right\}dq,
\ea 
\eeq 
where
\beq
\ba{ll}
F(q,t)&=\f{1}{4}L_1(t)\dot{L}_1(t)q^2+\f{1}{2}L_1(t)\dot{\a}(t)q+\ds\int_{0}^{t}\left[\f{\left(\dot{\a}^{2}(s)-4g_{0}(s)\right)L_1^2(s)-4\eps}{4L_{1}^{2}(s)}\right]ds.
\ea
\eeq 
Thus we have
\beq\label{average.Energy}
\bar{E}_n^{(\pm)}=h_{0,n}^{(\pm)}(t)+h_1(t)\,I_{1,n}^{(\pm)}+h_2(t)\,I_{2,n}^{(\pm)}, 
\eeq 
where
\beq
\ba{l}
h_{0,n}^{(\pm)}(t)=\ds-\f{\left(\dot{\a}^{2}(t)-4g_{0}(t)\right)L_1^2(t)-4E_n^{(\pm)}}{4L_{1}^{2}(t)}-i\left(\f{\dot{L}_1(t)}{2L_1(t)}+\f{\dot{\a}^2(t)}{2}\right),\\
h_1(t)=-\f{1}{2}\left(\dot{L}_1(t)\dot{\a}(t)+L_1(t)\ddot{\a}(t)\right)+i\left(\f{\dot{L}_1(t)}{2L_1(t)}-\dot{\a}(t)\dot{L}_1(t)\right),\\
h_2(t)=-\f{1}{4}\left(L_1(t)\ddot{L}_1(t)+\dot{L}_1^2(t)\right)-\f{i}{2}\dot{L}_1^2(t),\\
I_{k,n}^{(\pm)}=\ds\int \left[Q_n^{(\pm)}(q)\right]^2\,q^{k}\,dq,~k=1,2,\\
I_{3,n}^{(\pm)}=\ds\int \left(\f{\partial Q_n^{(\pm)}}{\partial q}\right)^2\, dq.
\ea
\eeq 
From the Eq. (\ref{average.Energy}) we can see that the average energy of the stationary state of a time dependent potential is complex in general whereas for time independent potential it is real. 

\section{Results and discussion}
\begin{figure}[H] 
	\centering
	\includegraphics[width=6cm,height=4cm]{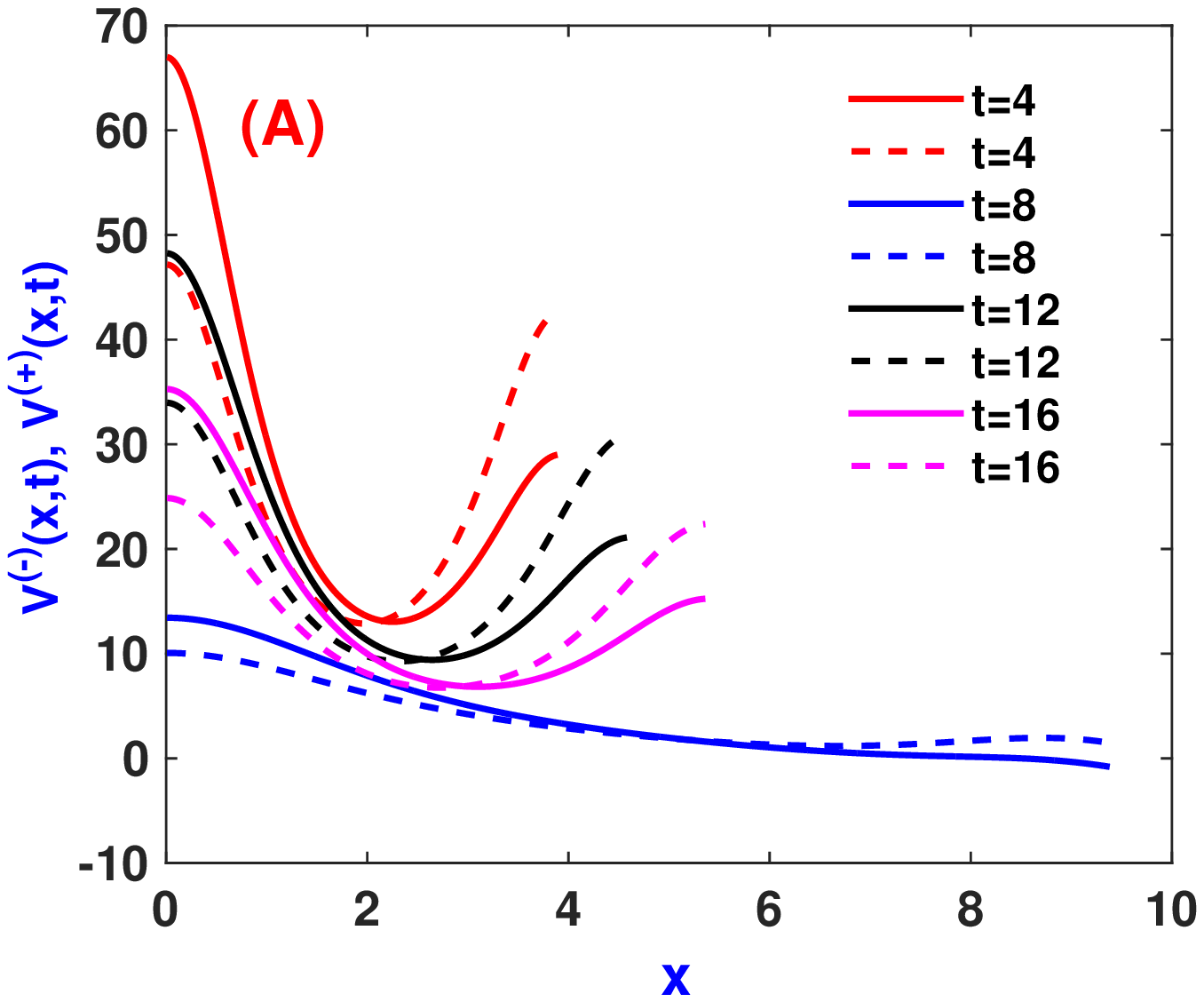}~~\includegraphics[width=6cm,height=4cm]{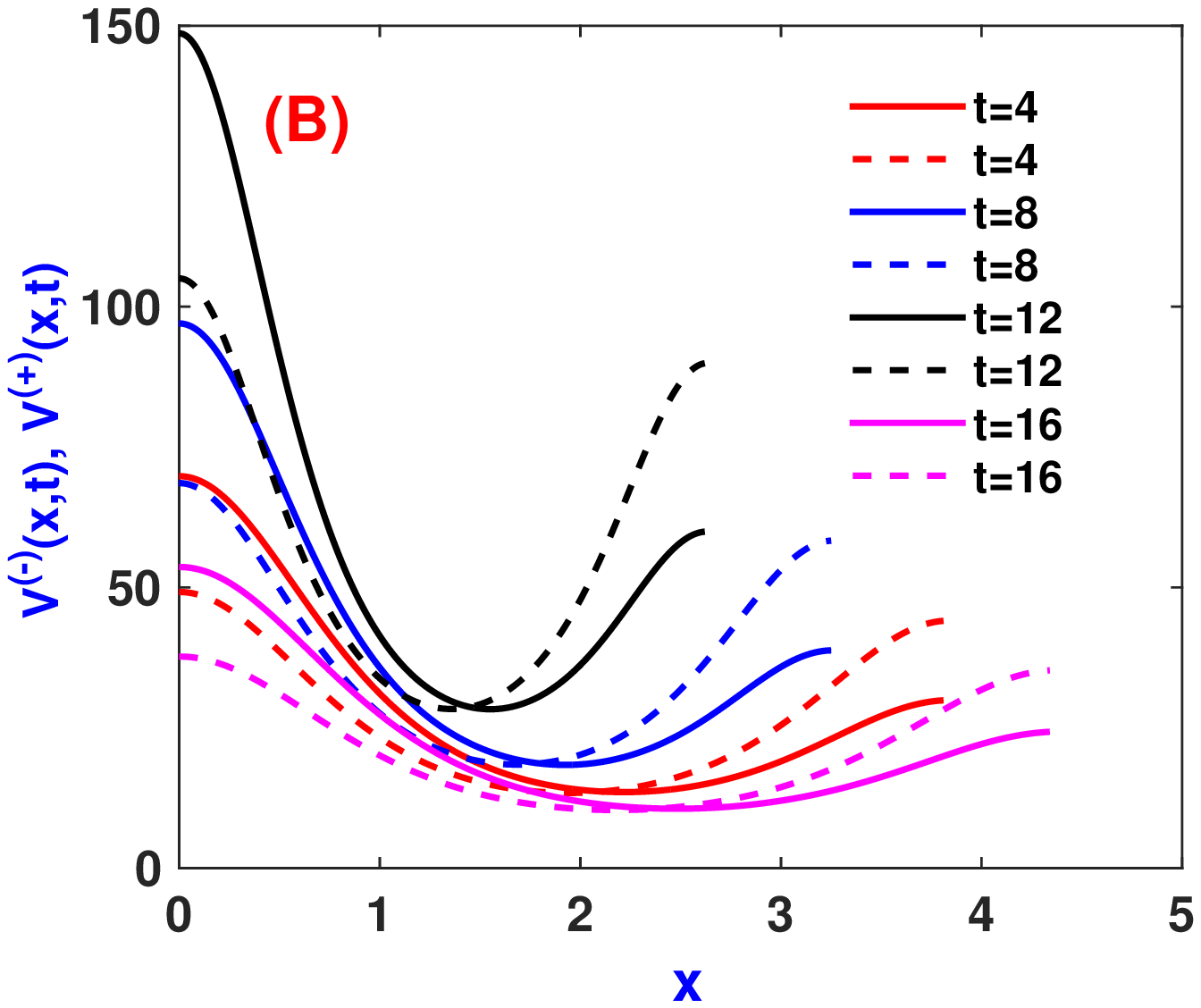}
	\caption{\label{Fig.CompPotxt} Plots of the potential $V^{(-)}(x,t)$ (solid curve) with $V^{(+)}(x,t)$ (dashed curve), for (A) $L(t)=\pi(2+\sin t)$ and (B) $L(t)=\f{A_1\pi}{\sqrt{1+B_1\cos\om t}}$ and $A,=5,B=0.2,a=\sqrt{\frac{2}{3}},b=\sqrt{2},\om=1$.}
\end{figure}

So far our analysis has been quite general in the sense that no specific form of the quantity $L(t)$ was necessary. However, to obtain quantitative understanding of the results it now becomes necessary to prescribe specific form(s) of $L(t)$. Here we shall consider the following forms of $L(t)$:
\beq
\ba{l}
L(t)=L^{(1)}(t)=\ds\pi(2+\sin t),\\
L(t)=L^{(2)}(t)=\ds\f{A_1\pi}{\sqrt{1+B_1\cos\om t}},
\ea
\eeq  
where $A_1=\f{ab\sqrt{2}}{\sqrt{a^2+b^2}}$, $B_1=\f{b^2-a^2}{b^2+a^2}$ are free parameters. To get a visual understanding of how the time dependent potential look at different points of time, we have presented in Fig \ref{Fig.CompPotxt} plots of the potentials $V^{\pm}(x,t)$ for $L(t)=\pi(1+\sin t),~ \f{A_1\pi}{\sqrt{1+B_1\cos\om t}}$ for different values of $t$. It is seen from Fig \ref{Fig.CompPotxt} that both the potentials show well structure although the wells for $V^+(x,t)$ are deeper than those of $V^-(x,t)$ for each value of $t$. This is true for both the choices of $L(t)$.

\begin{figure}[H] 
	\centering
	\includegraphics[width=6cm,height=4cm]{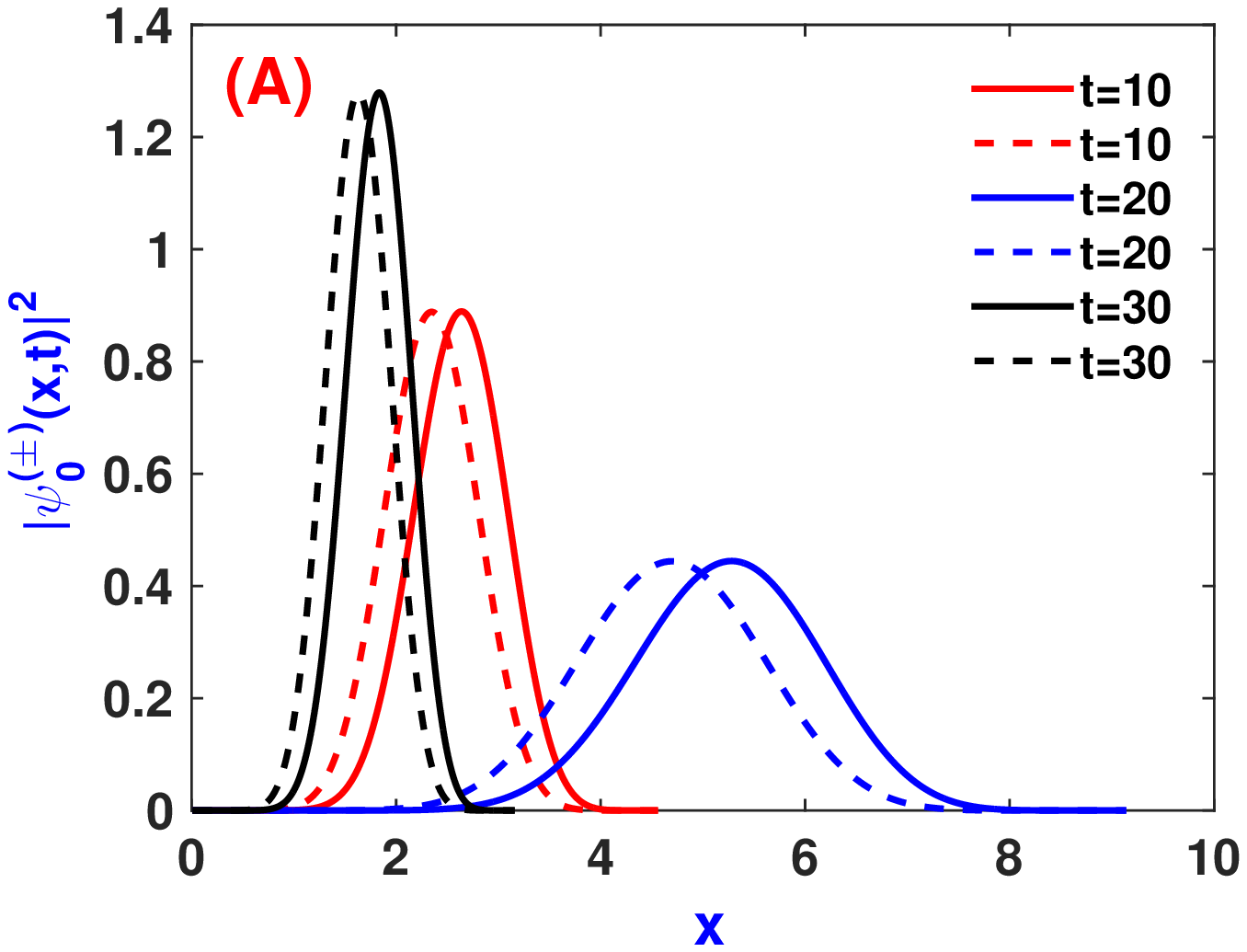}~~\includegraphics[width=6cm,height=4cm]{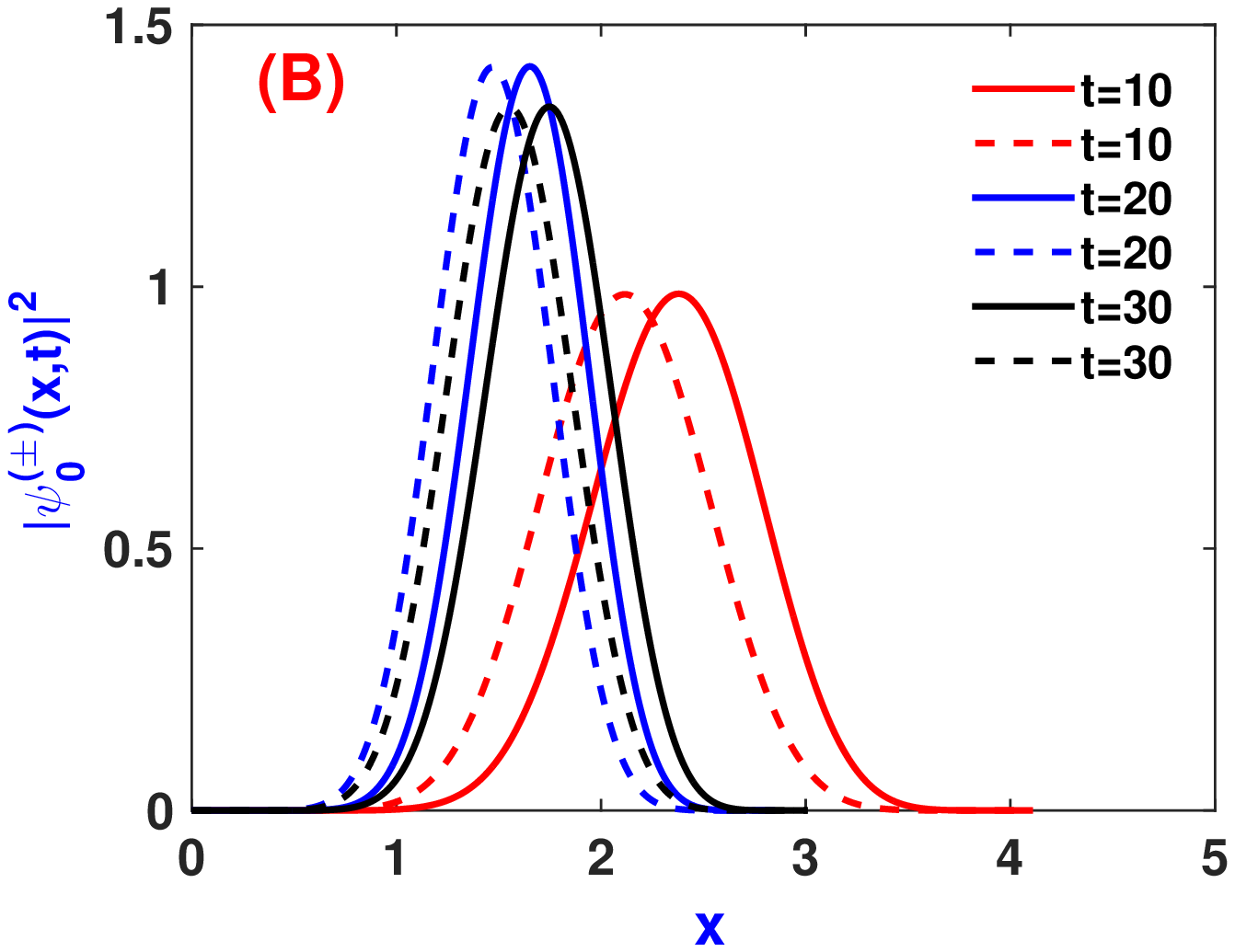}\\
	\includegraphics[width=6cm,height=4cm]{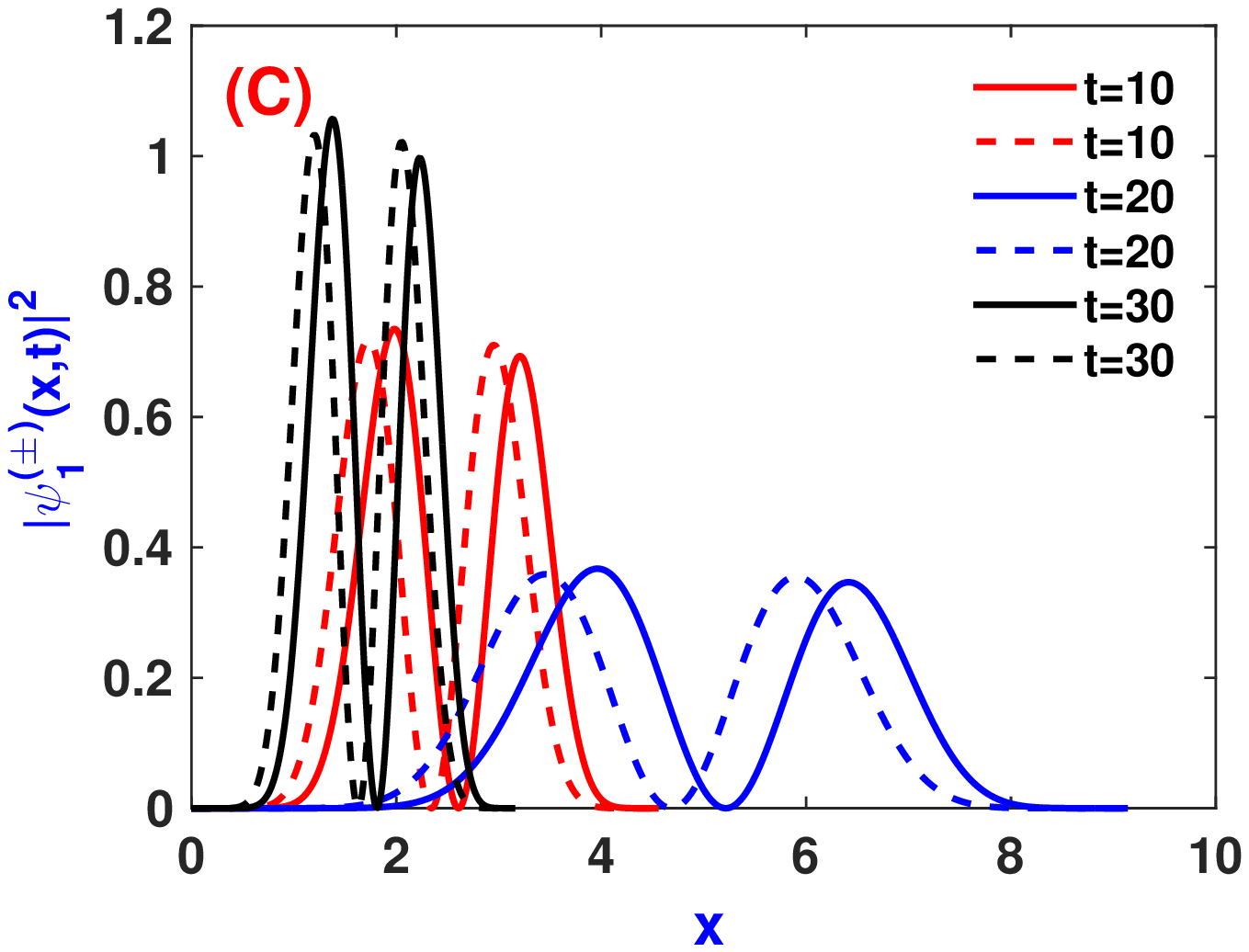}~~\includegraphics[width=6cm,height=4cm]{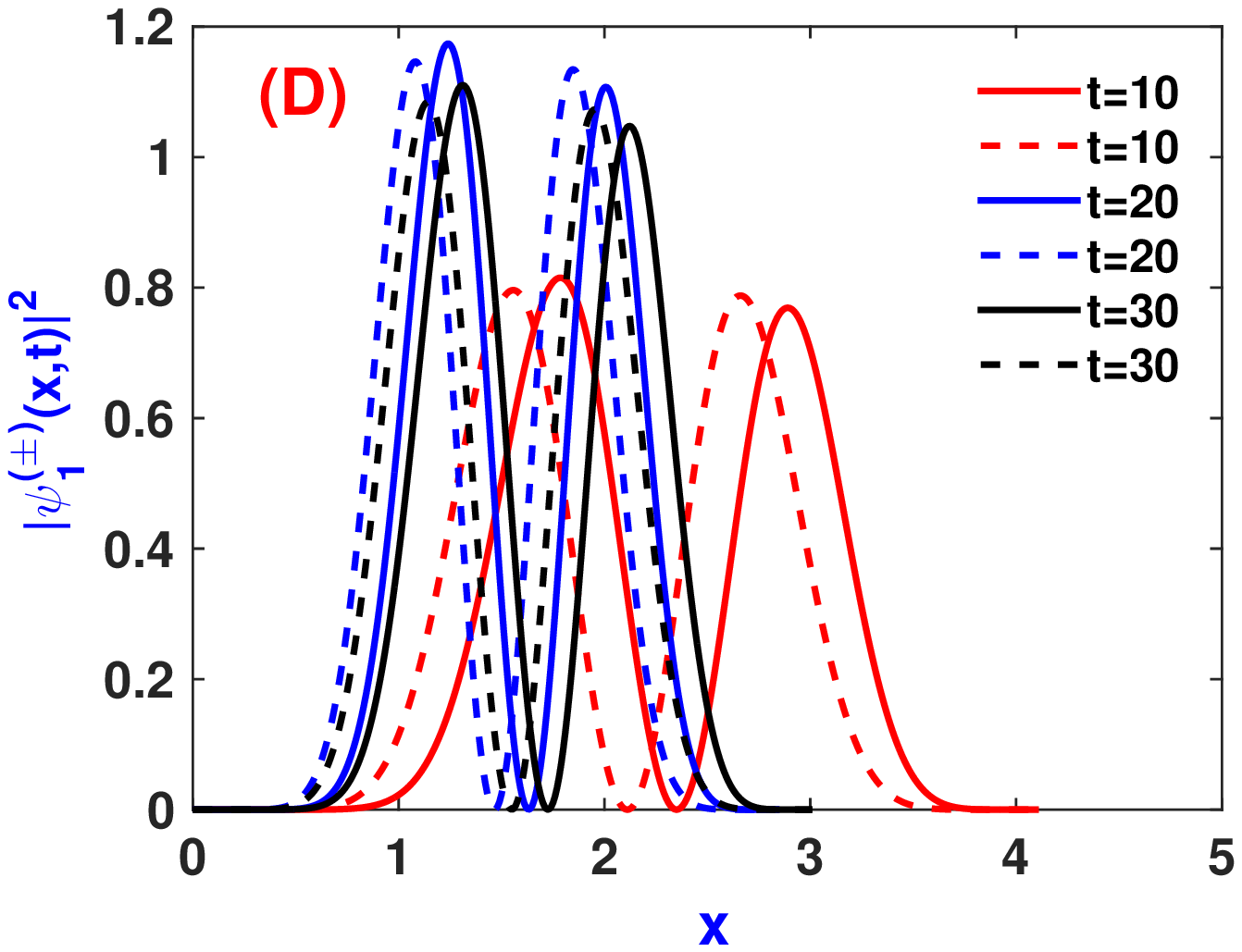}\\
	\includegraphics[width=6cm,height=4cm]{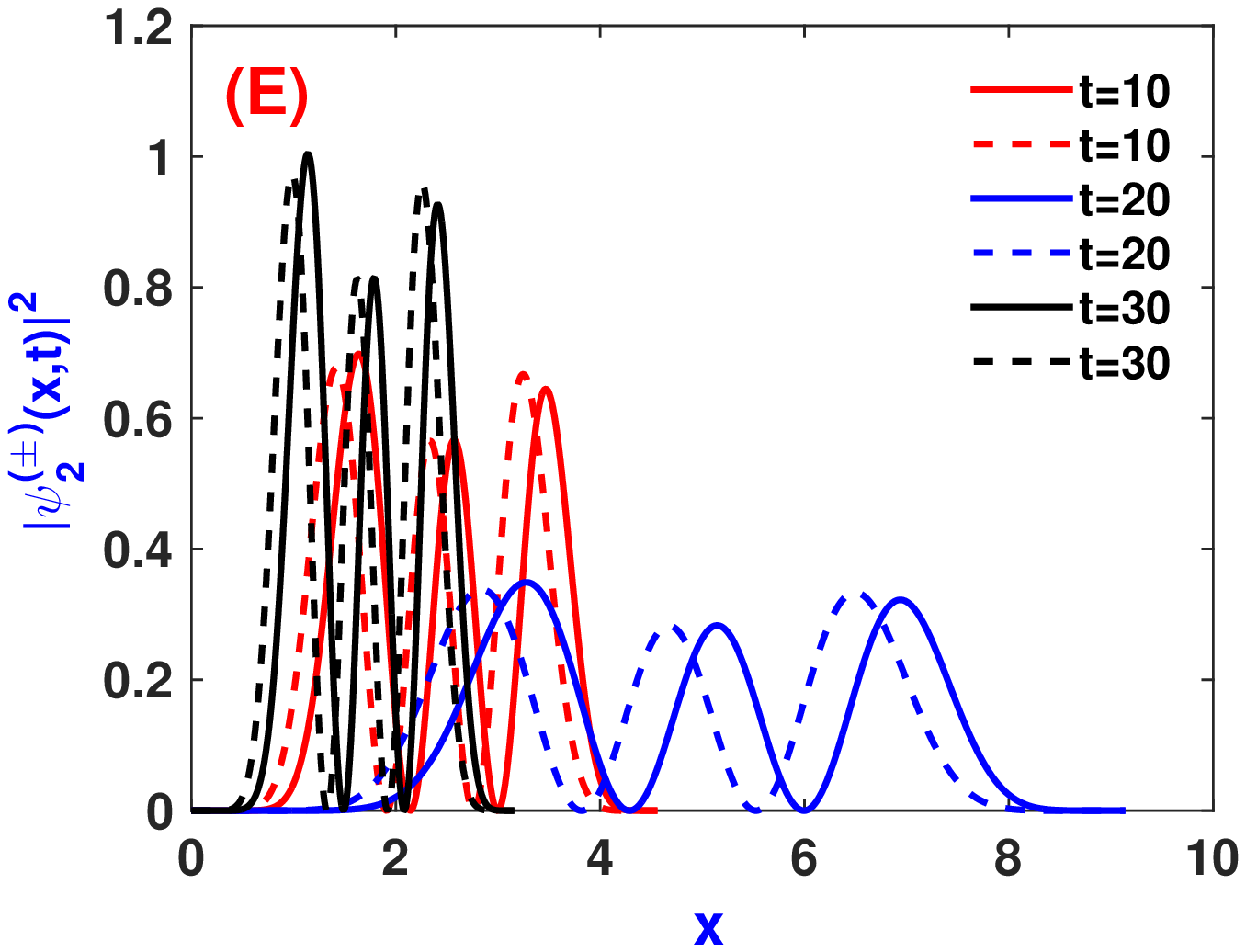}~~\includegraphics[width=6cm,height=4cm]{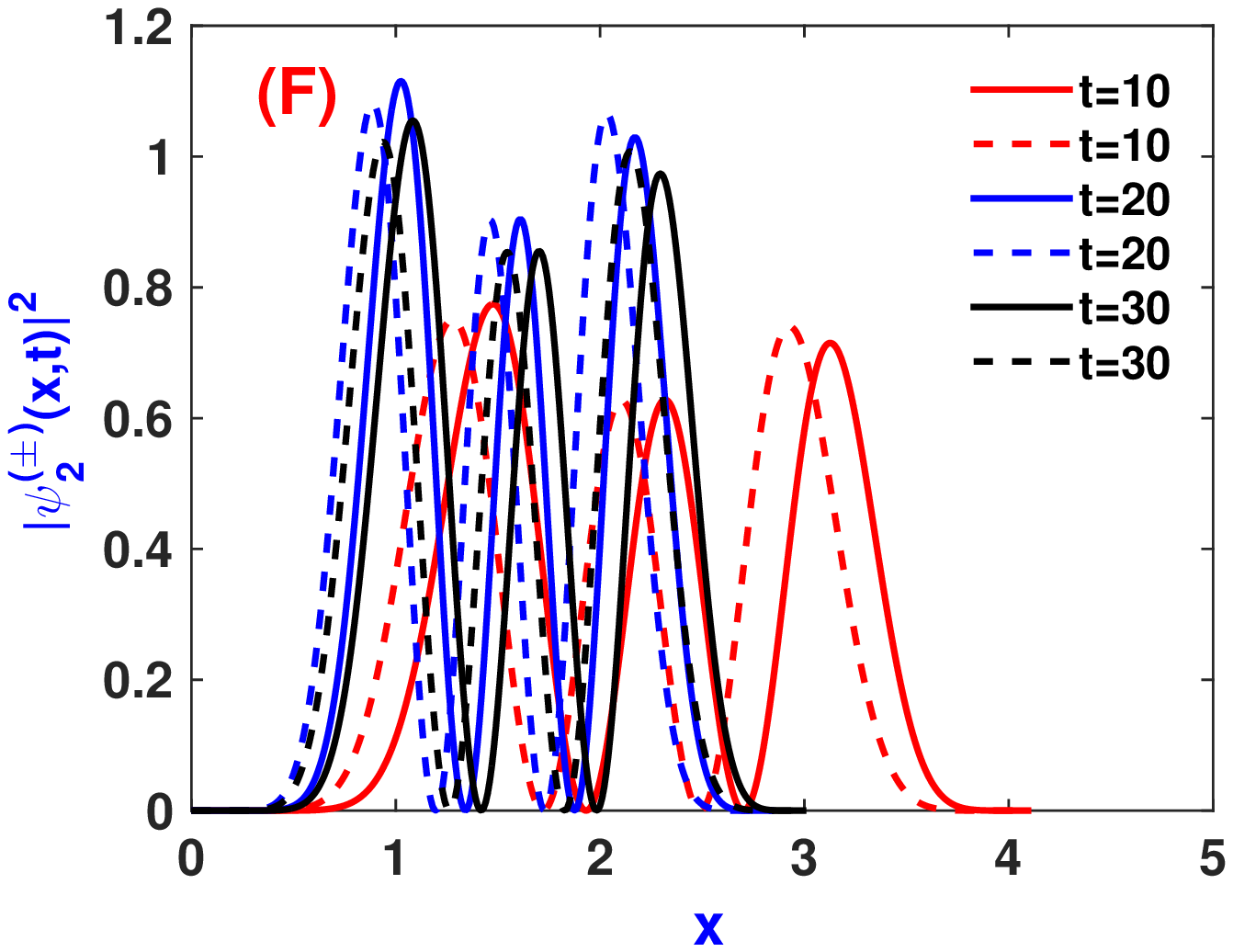}
	\caption{\label{Fig.Comp.densities} Plots of probability densities $\rho_n^{(-)}(x,t)$ (solid curve) with $\rho_n^{(+)}(x,t)$ (dashed curve), for (A) n=0, (B) n=0, (C) n=1, (D) n=1, (E) n=2, (F) n=2. The left column is defined for $L(t)=\pi(2+\sin t)$ whereas the right column is defined for $L(t)=\f{A_1\pi}{\sqrt{1+B_1\cos\om t}}$. The other parameters are taken as $A,=5,B=0.2,a=\sqrt{\frac{2}{3}},b=\sqrt{2},\om=1$.}
\end{figure}

To avoid the singularity we consider $A>Max\left\{B+1.5,|B|+0.5\right\}$. Next, we examine the instantaneous probability densities $\rho_n^{\pm}=|\psi_n^{\pm}(x,t)|^2$ at different times. A plot of the instantaneous densities are given in Fig \ref{Fig.Comp.densities}. From the figure we see that particles in both the potentials have more or less similar localization behavior at different values of time $t$ although the wells are narrower in the $(+)$ sector. It may also be observed that localization is periodic in nature. For example, starting from $t=10$, localization increases as $t$ increases to $t=20$ and thereafter it decreases as $t$ increases to $t=30$. This is generally true for all the levels. 

\begin{figure}[H] 
	\centering
	\includegraphics[width=6cm,height=4cm]{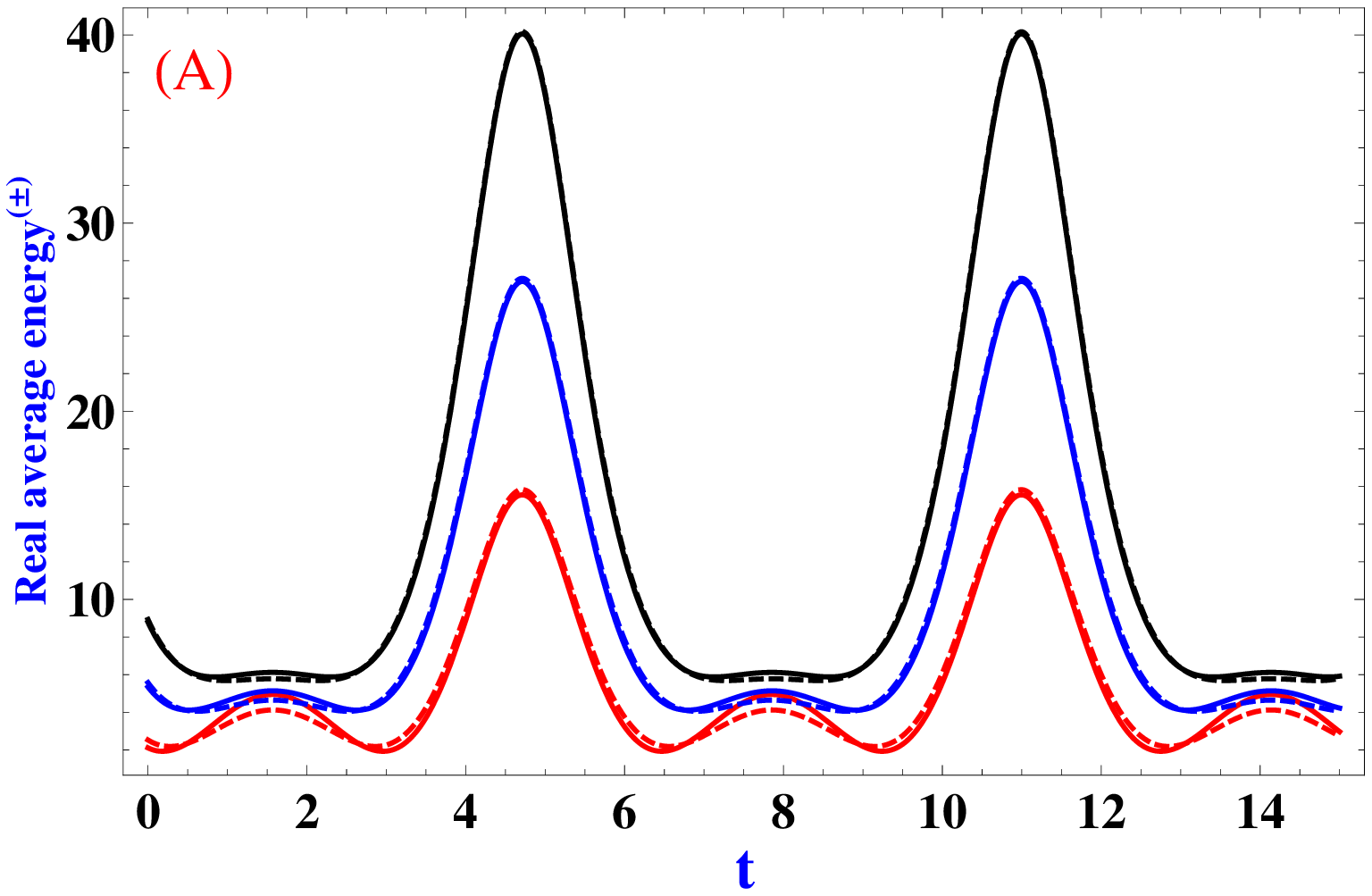}~\includegraphics[width=6cm,height=4cm]{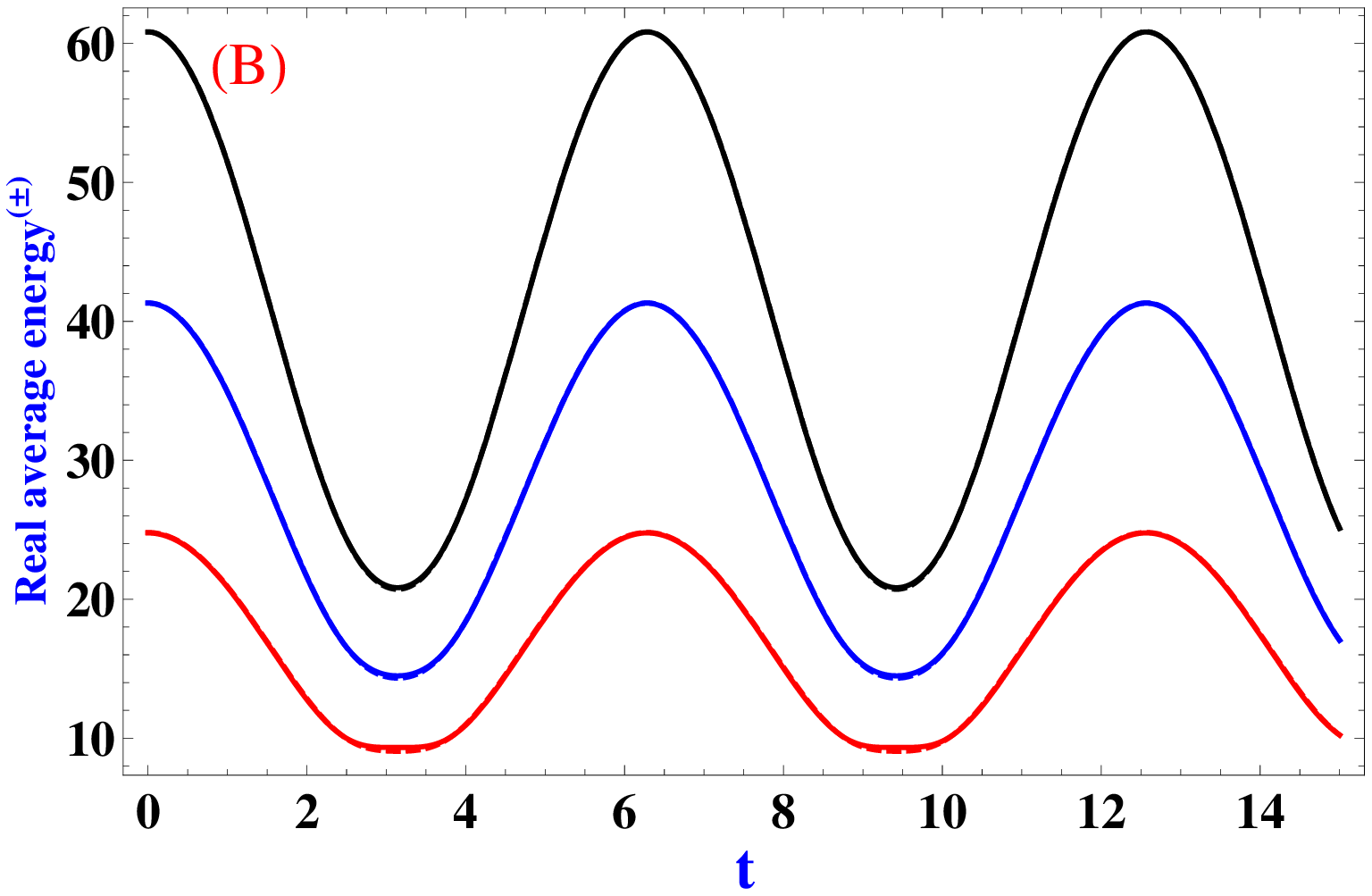}\\
	\includegraphics[width=6cm,height=4cm]{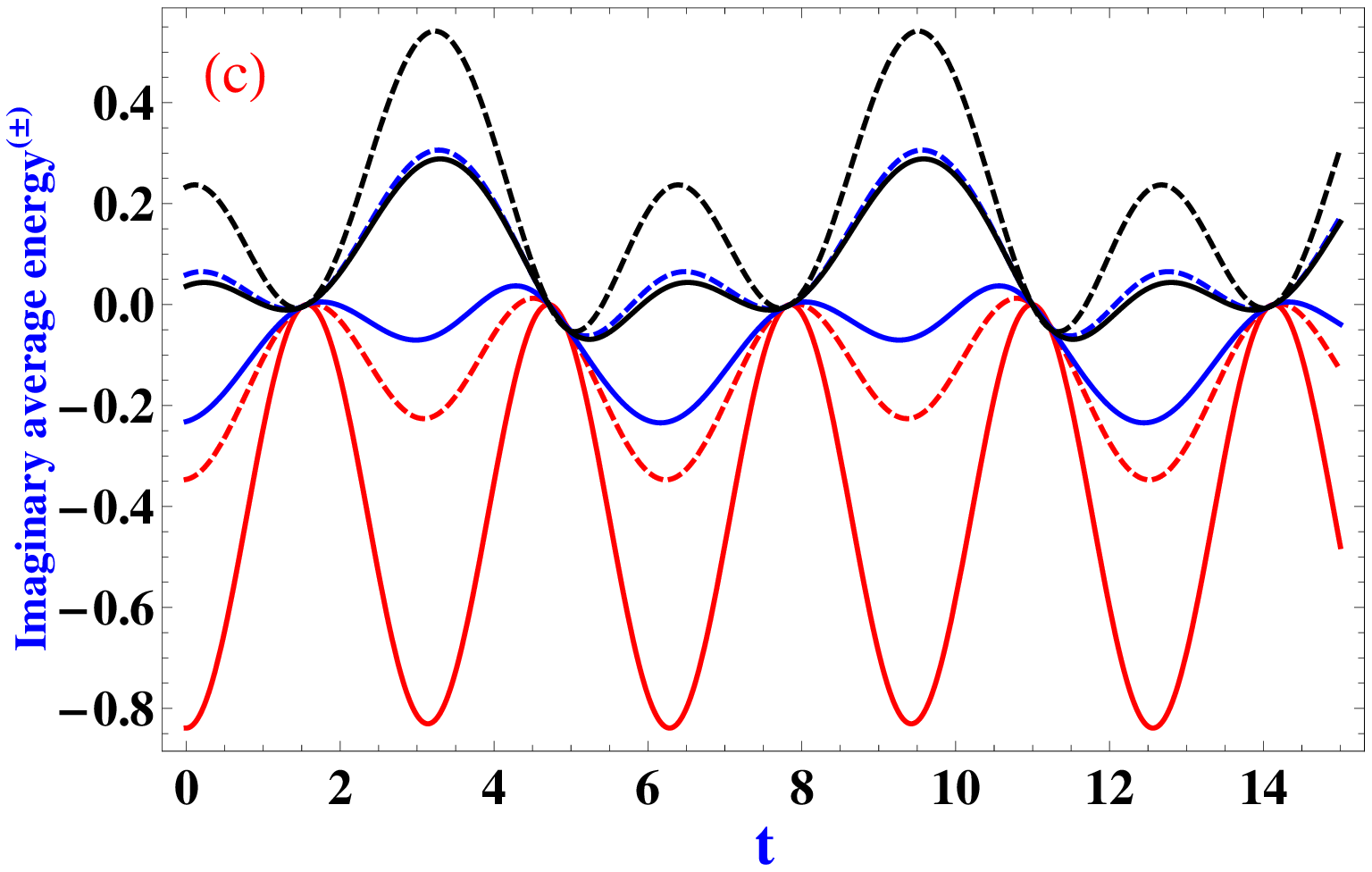}~\includegraphics[width=6cm,height=4cm]{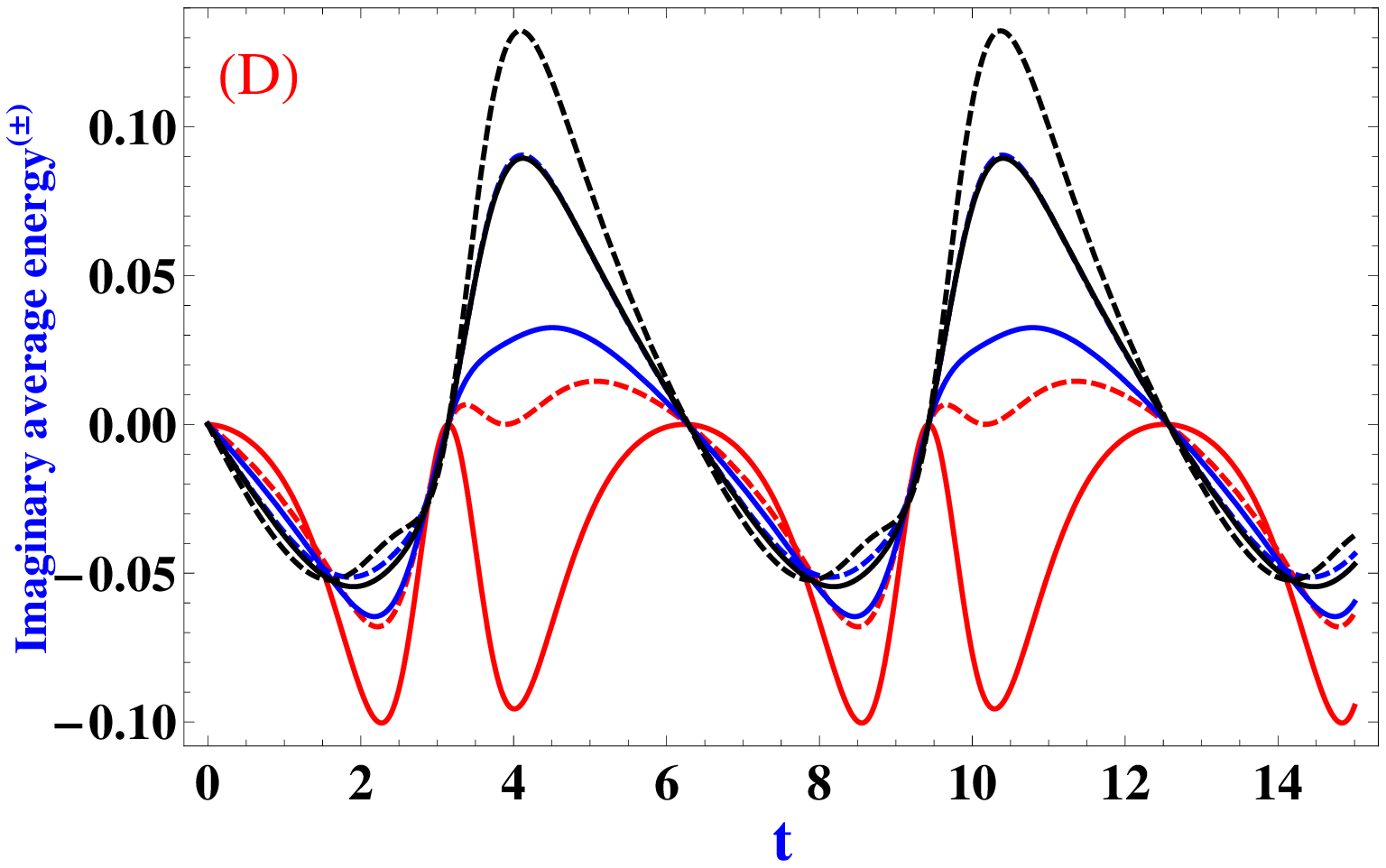}
	\caption{\label{Fig.Compare.aveEenergy} Comparison of the average energy: (1) (A)-(B): real parts of $\left(\bar{E}_{0}^{(-)},\bar{E}_{0}^{(+)}\right)$ (red), $\left(\bar{E}_{1}^{(-)},\bar{E}_{1}^{(+)}\right)$ (blue) and $\left(\bar{E}_{2}^{(-)},\bar{E}_{2}^{(+)}\right)$ (black), (2) (C)-(D): imaginary part of $\left(\bar{E}_{0}^{(-)},\bar{E}_{0}^{(+)}\right)$ (red), $\left(\bar{E}_{1}^{(-)},\bar{E}_{1}^{(+)}\right)$ (blue) and $\left(\bar{E}_{2}^{(-)},\bar{E}_{2}^{(+)}\right)$ (black). The solid curves denote for $\psi_{0,1,2}^{(-)}$ and the dashed curves denote for $\psi_{0,1,2}^{(+)}$. In the left column we have taken $L(t)=\pi(2+\sin t)$ while in the right column $L(t)=\f{A_1\pi}{\sqrt{1+B_1\cos\om t}}$. The other parameters are $A=5, B=3.4, A_1=1; B_1=0.5, \omega=1$.}
\end{figure}

\begin{figure}[H] 
	\centering
	\includegraphics[width=6cm,height=4cm]{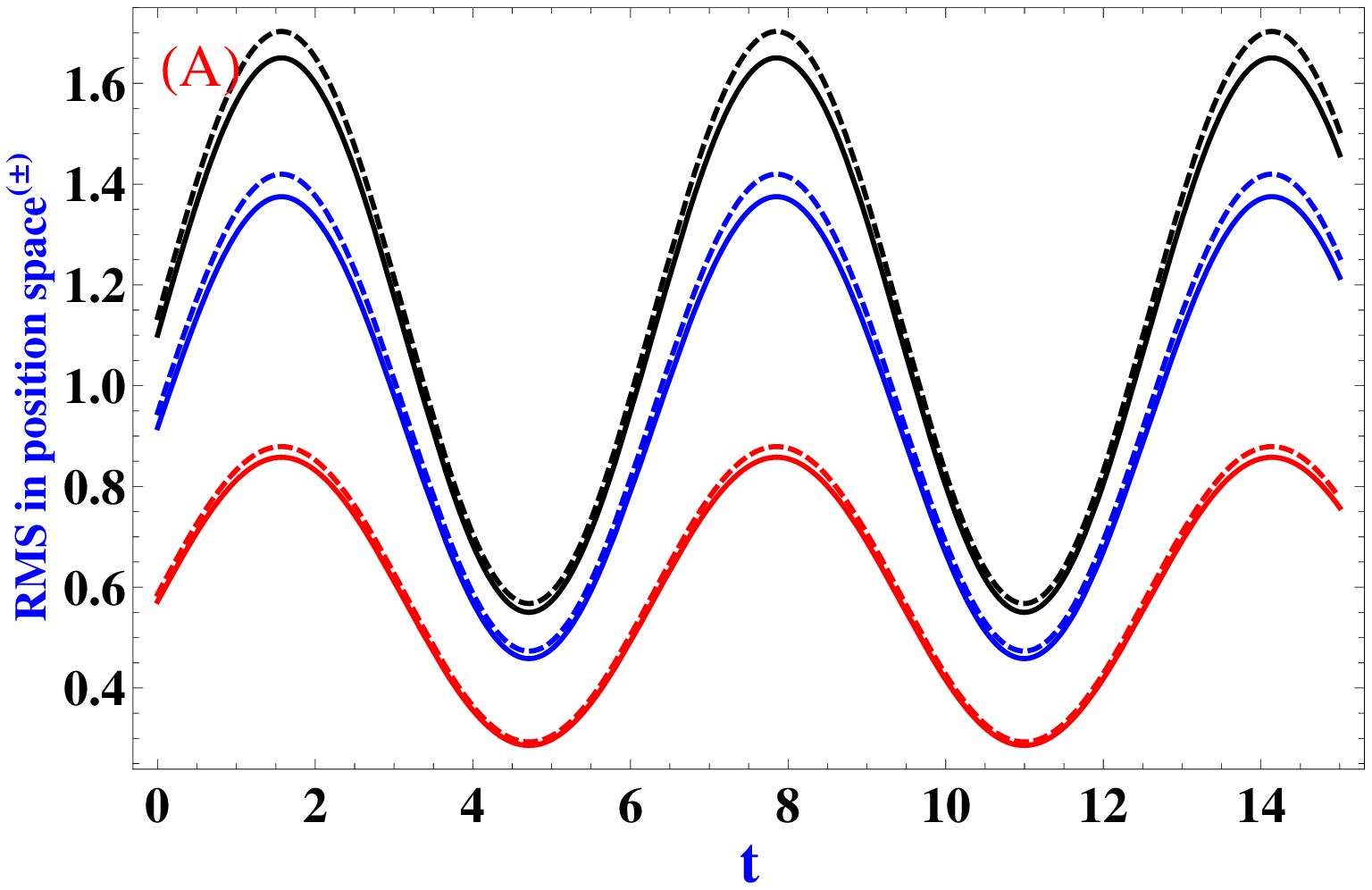}~\includegraphics[width=6cm,height=4cm]{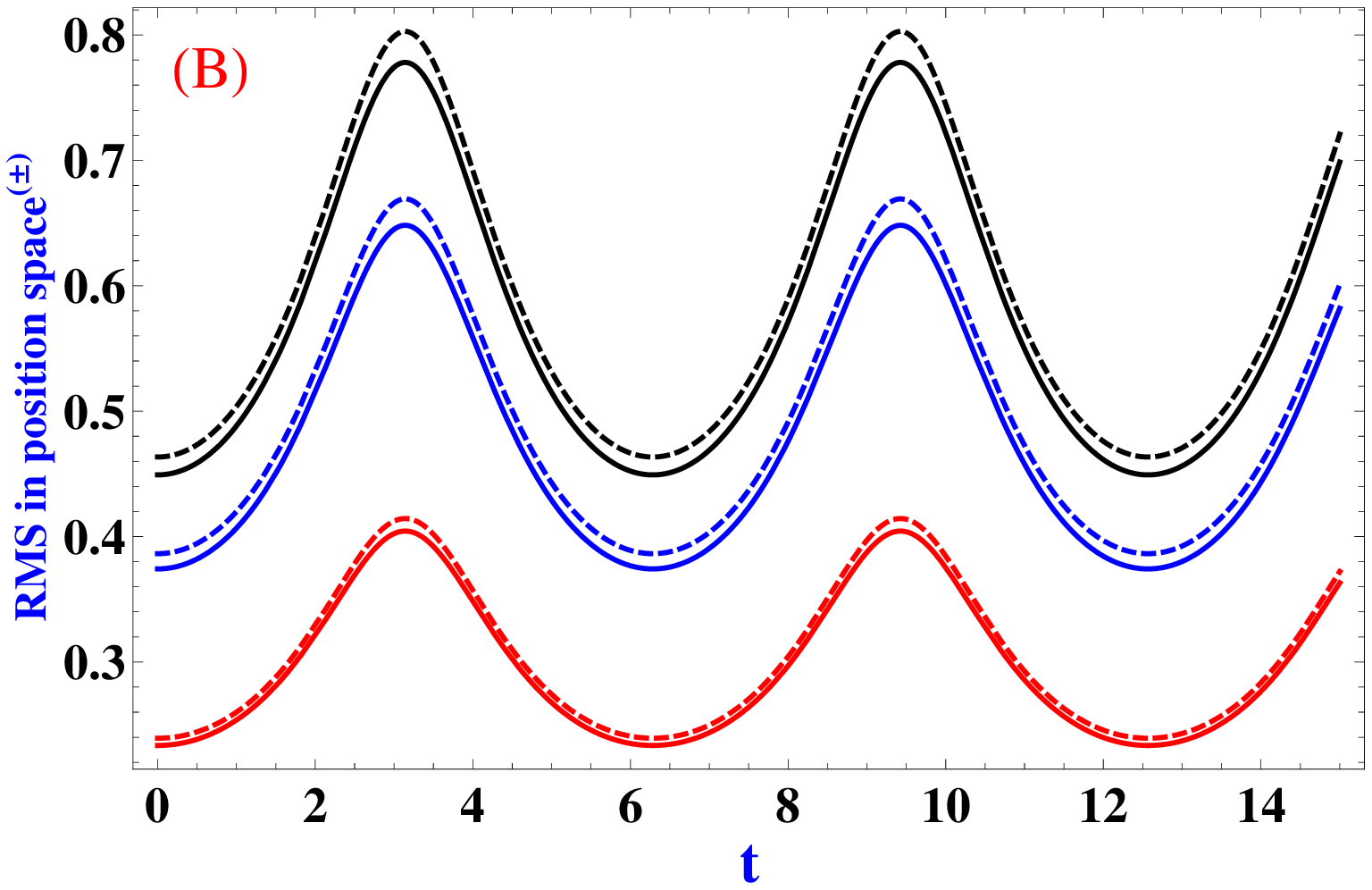}\\
	\includegraphics[width=6cm,height=4cm]{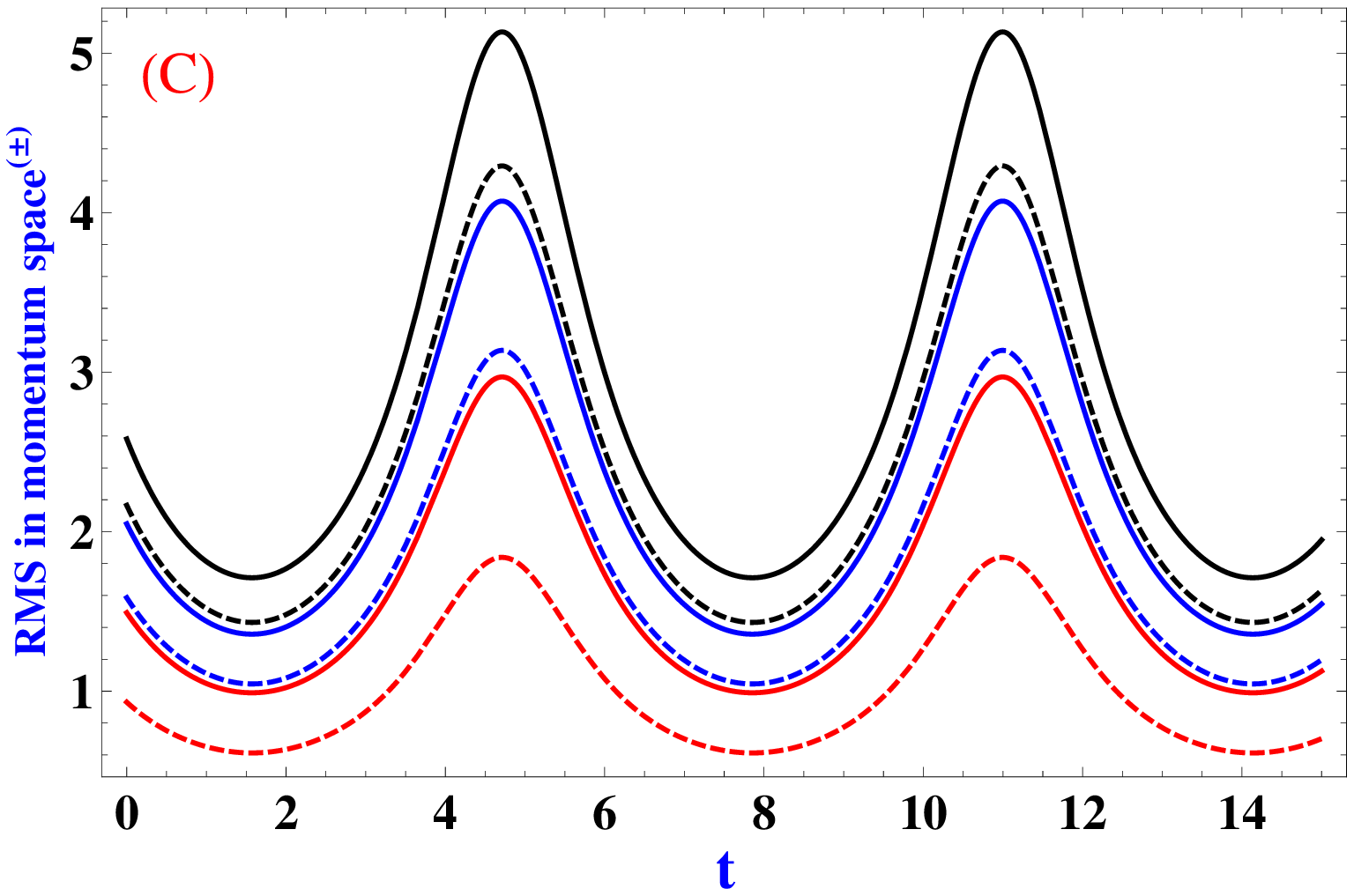}~\includegraphics[width=6cm,height=4cm]{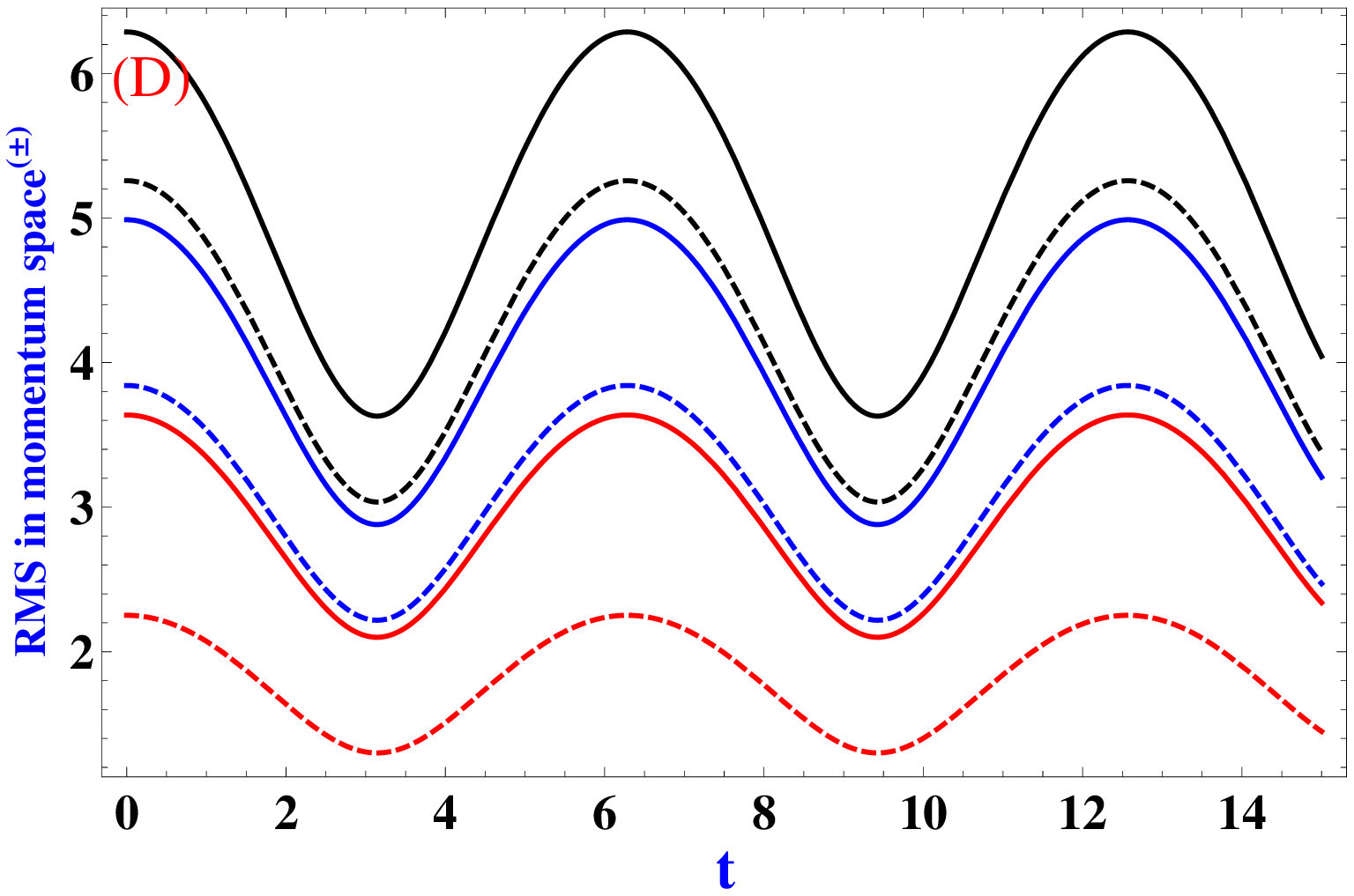}
	\caption{\label{Fig.Compare.std.xp} Comparison of the RMS in position space (A)-(B) $\left(\left(\Delta x\right)_{0}^{(-)}, \left(\Delta x\right)_{0}^{(+)}\right)$ (red),  $\left(\left(\Delta x\right)_{1}^{(-)},\left(\Delta x\right)_{1}^{(+)}\right)$ (blue) and  $\left(\left(\Delta x\right)_{2}^{(-)},\left(\Delta x\right)_{2}^{(+)}\right)$ (black), in momentum space (C)-(D) $\left(\left(\Delta p\right)_{0}^{(-)}, \left(\Delta p\right)_{0}^{(+)}\right)$ (red),  $\left(\left(\Delta p\right)_{1}^{(-)},\left(\Delta p\right)_{1}^{(+)}\right)$ (blue) and  $\left(\left(\Delta p\right)_{2}^{(-)},\left(\Delta p\right)_{2}^{(+)}\right)$ (black). The solid curves represent the $(-)$ sector while dashed curves represent $(+)$ sector. In the left panel $L(t)=\pi(2+\sin t)$ and in the right panel $L(t)=\f{A_1\pi}{\sqrt{1+B_1\cos\om t}}$. The parameter values are $A=5, B=3.4, A_1=1; B_1=0.5, \omega=1$.}
\end{figure}

We now examine the behavior of the average energy ${\bar E}$. It can be seen that the average energy depends on the first and second order moments of the variable $q$ of the time independent Schr\"odinger equation. It is however difficult to obtain analytically the first order moment $I_{1,n}^{(\pm)}$ and second order moment $I_{2,n}^{(\pm)}$ of $q$ for the states $Q_0^{(\pm)}$. Therefore we have evaluated numerically the moments $I_{1,n}^{(\pm}$ and $I_{2,n}^{(\pm)}$ for $n=0,1,2$. In Fig \ref{Fig.Compare.aveEenergy} we have presented plots of the real and imaginary parts of the average energy corresponding to both the potentials $V^{\pm}(x,t)$. It can be seen from (A)-(B) of Fig \ref{Fig.Compare.aveEenergy} that the real parts of the average energy exhibits similar behavior and are almost identical at all times while from (C)-(D) we find that the imaginary parts of the average energy coincide at particular instants of time and for other values of $t$ they are different for both the $(\pm)$ sectors. In addition it is to be noted that the average energies $\bar{E}_n^{(\pm)}$ are equivalent to energies $\f{\pi^2}{L^2}\,E_n^{(\pm)}$ of the time independent Schr\"odinger equation when the boundary conditions are fixed, where $E_n^{(\pm)}$ are the energies of the effective potentials $\widetilde{V}^{(\pm)}$.
Now we discuss the RMS. In Fig \ref{Fig.Compare.std.xp} we have presented plots of the RMS. From the figure we find that the RMS of the $(+)$ sector is less than the same of the $(-)$ sector for all values of $n$. This can be attributed to the fact that solutions of the $(+)$ sector are given in terms exceptional orthogonal polynomials. Also in both the sectors RMS is smaller for smaller values of $n$. Similar behavior can be seen in the momentum space also. Although the RMS are quantitatively different for the two profiles of $L(t)$, they are qualitatively the same for both the profiles.

\begin{figure}[H]  
	\centering
	\includegraphics[width=6cm,height=4cm]{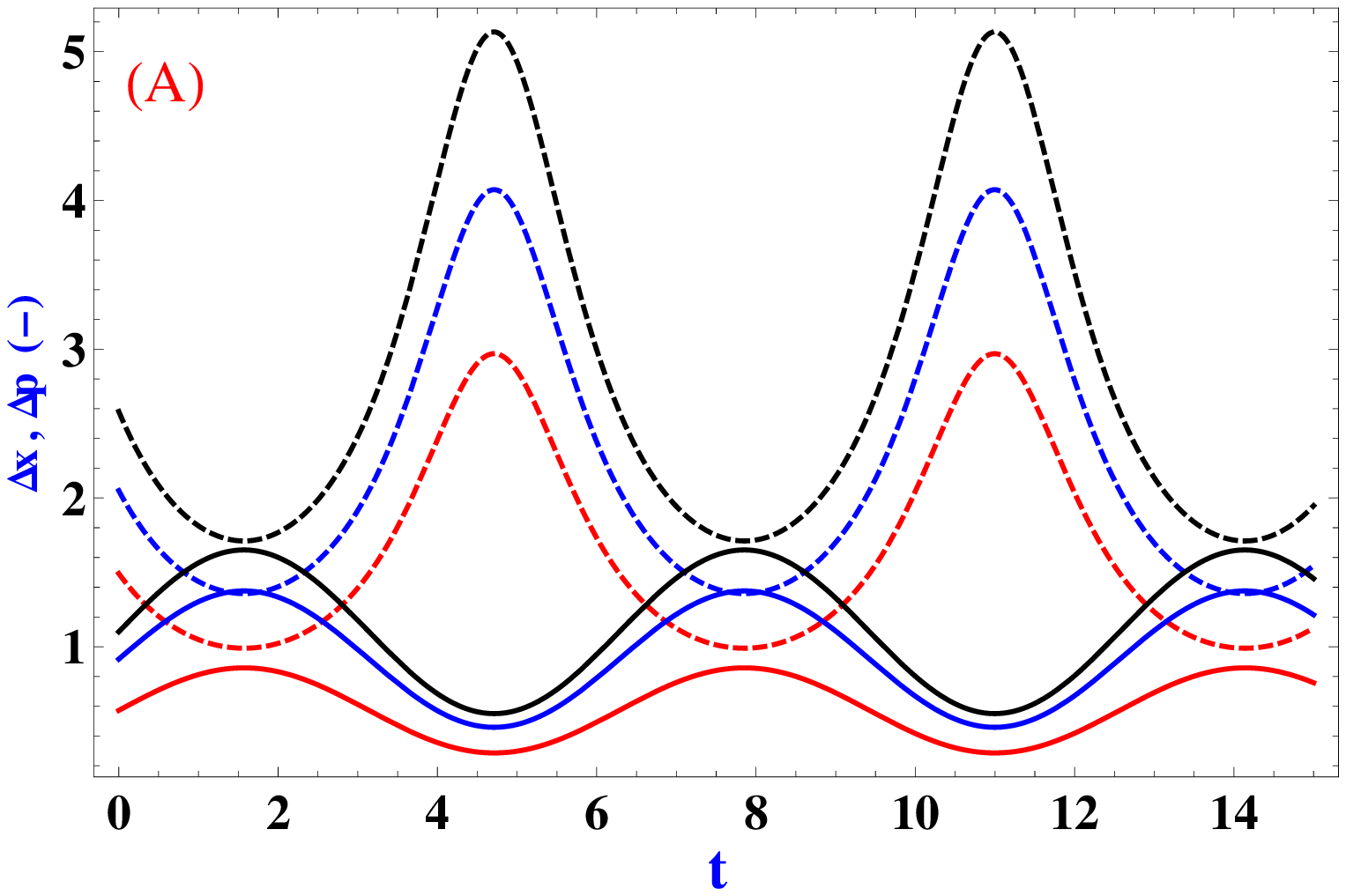}~~\includegraphics[width=6cm,height=4cm]{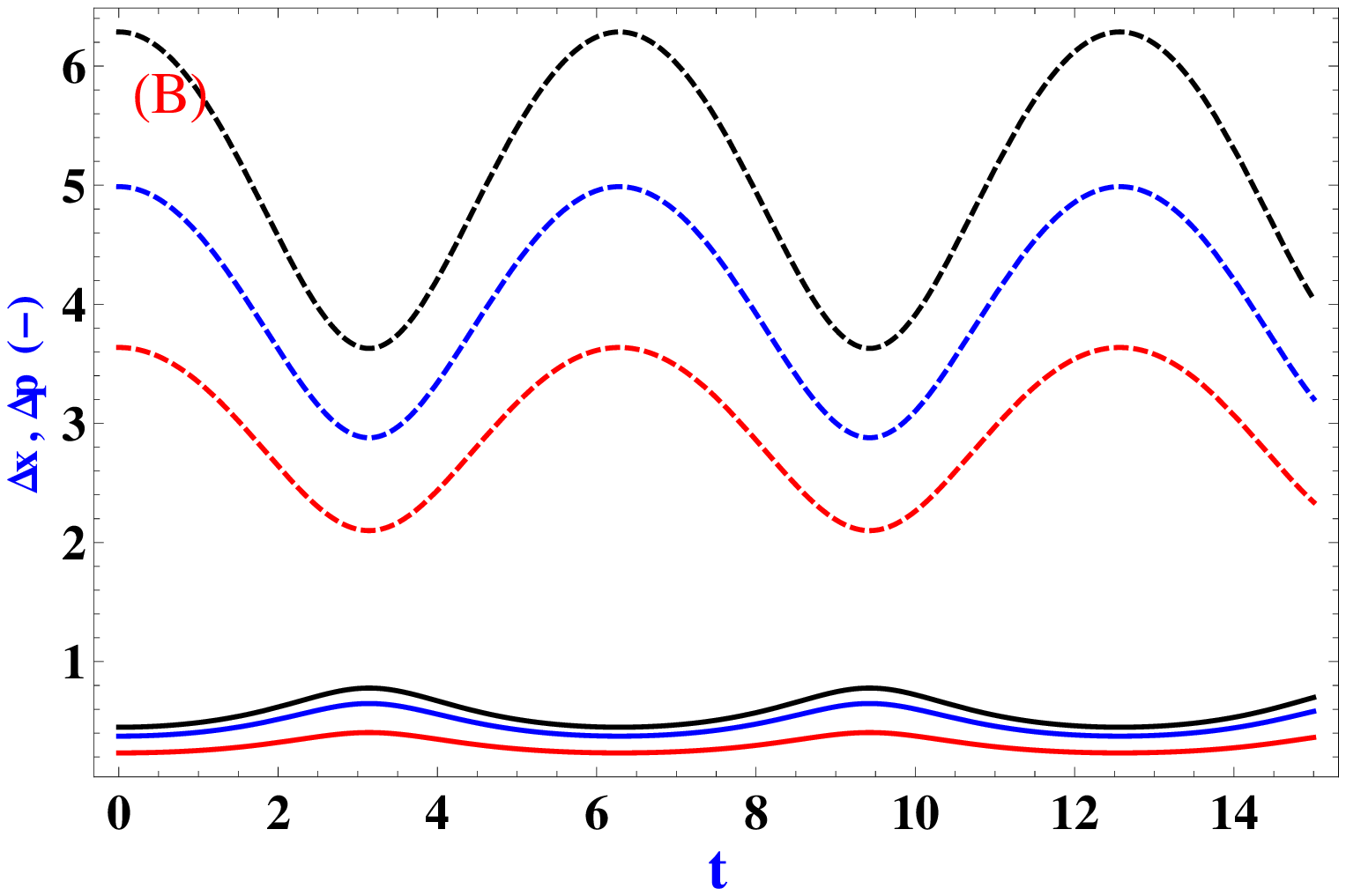}\\
	\includegraphics[width=6cm,height=4cm]{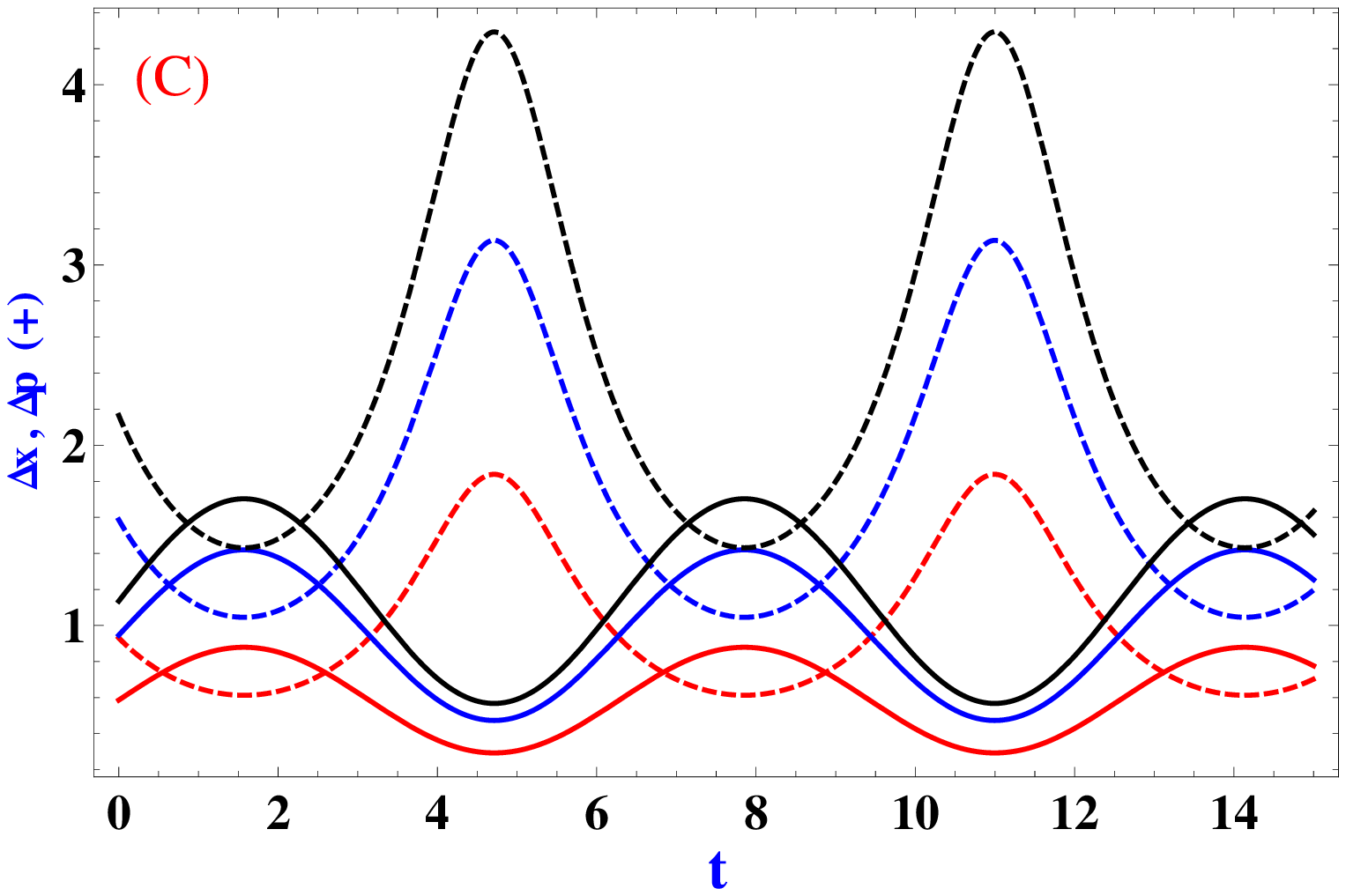}~~\includegraphics[width=6cm,height=4cm]{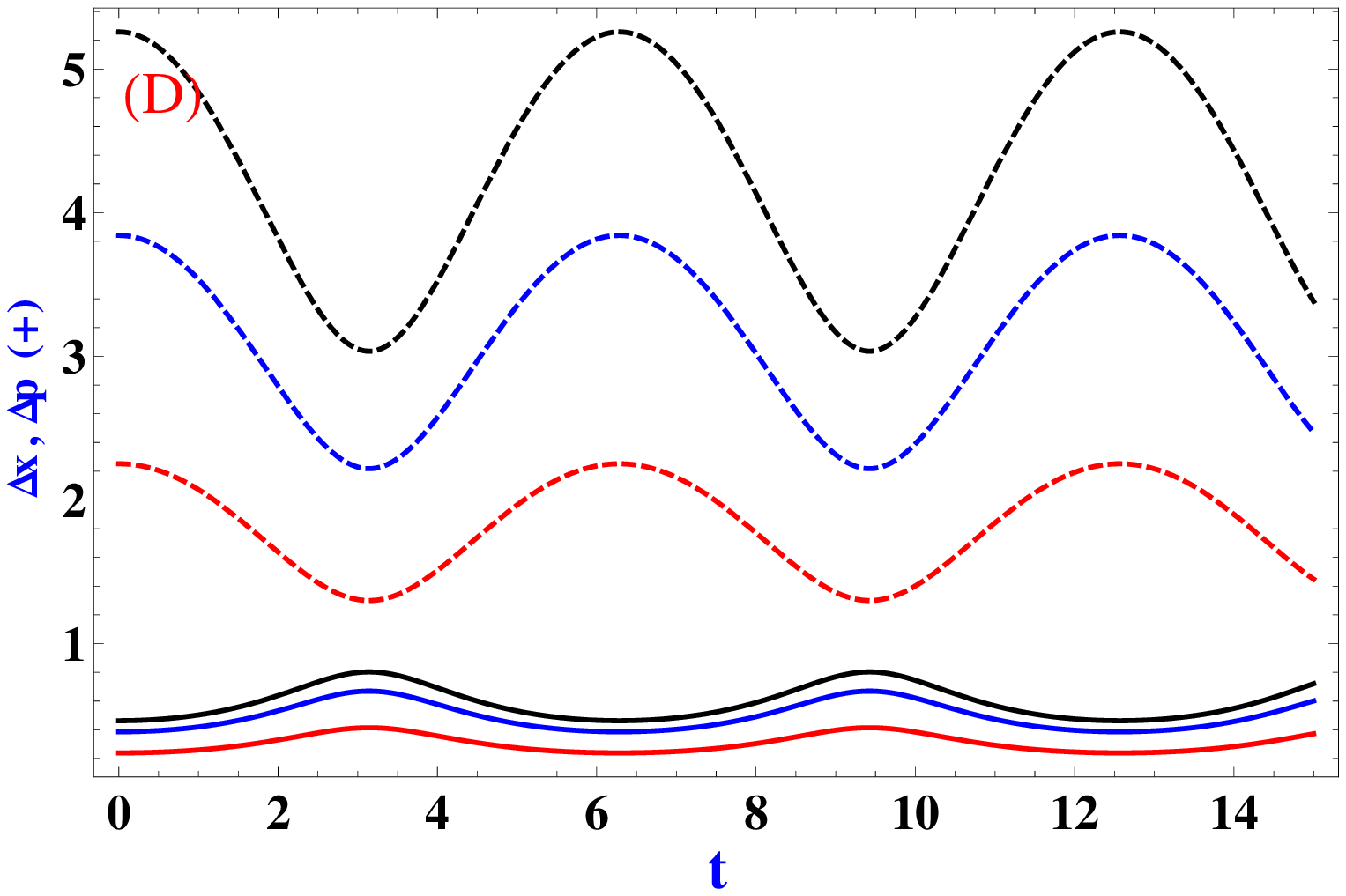}
	\caption{\label{Figdelxdelp} Variation of the uncertainty in position space (solid curves) and  momentum space (dashed curves) against time. The left column is defined for $L(t)=\pi(2+\sin t)$, and the right column is defined for $L(t)=\f{A_1\pi}{\sqrt{1+B_1\cos\om t}}$ with red $(n=0)$, blue $(n=1)$ and black $(n=2)$. The parameter values are $A=5, B=3.4, A_1=1; B_1=0.5, \omega=1$.}
\end{figure}

It may be pointed out that for the uncertainty relation to be satisfied, the behavior of the RMS $(\Delta x)_n^\pm$ and $(\Delta p)_n^\pm$ should be opposite for all values of $t$. In other words, if one of the quantities is increasing the other must be decreasing and vice-versa. In Fig \ref{Figdelxdelp}  we have presented plots of $(\Delta x)_n^\pm$ and $(\Delta p)_n^\pm$ against $t$ for both the choices of $L(t)$. The figure shows the desired features thus confirming the above assertion.

\begin{figure}[H] 
	\centering
\includegraphics[width=6cm,height=4cm]{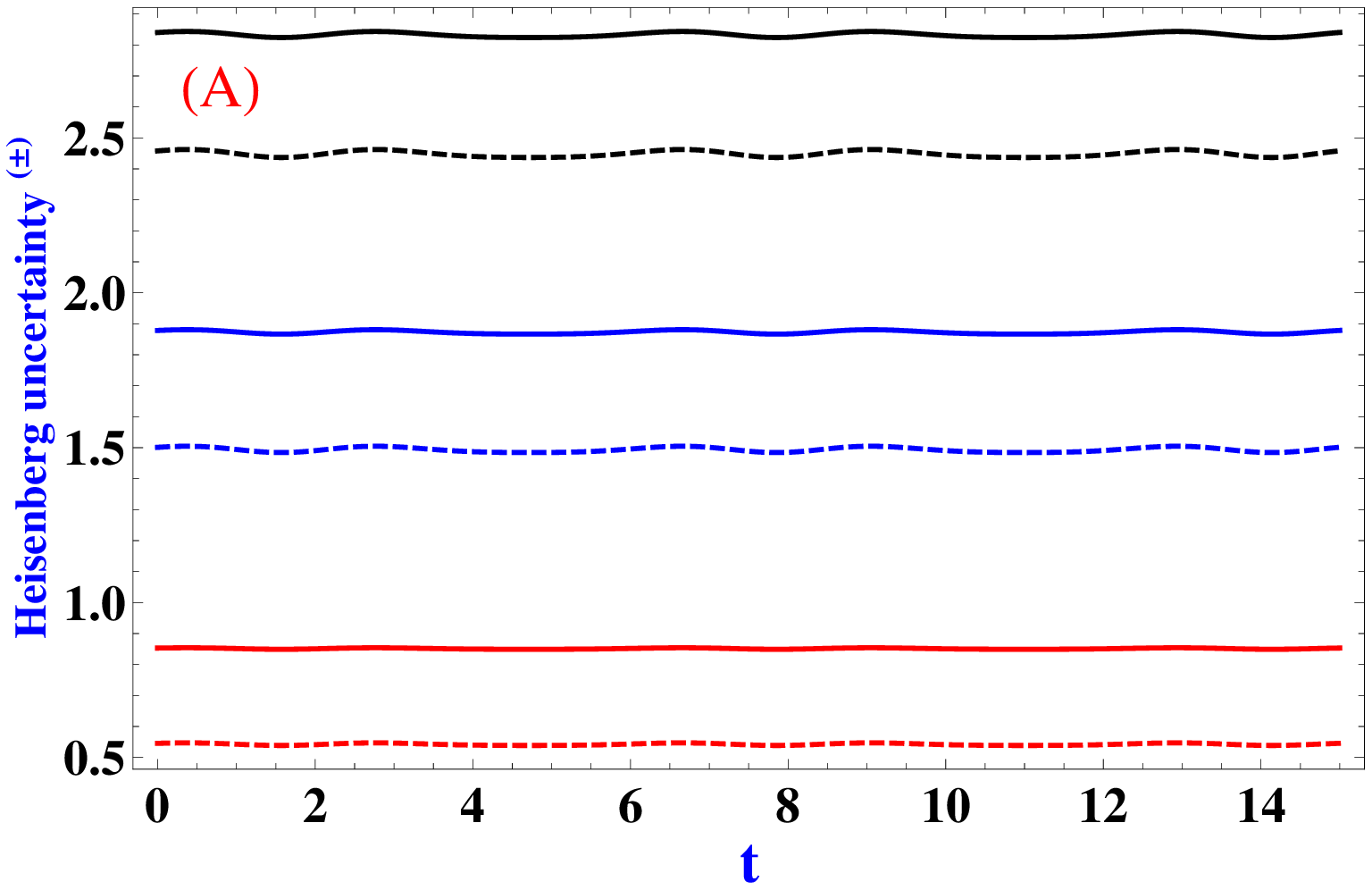}~~\includegraphics[width=6cm,height=4cm]{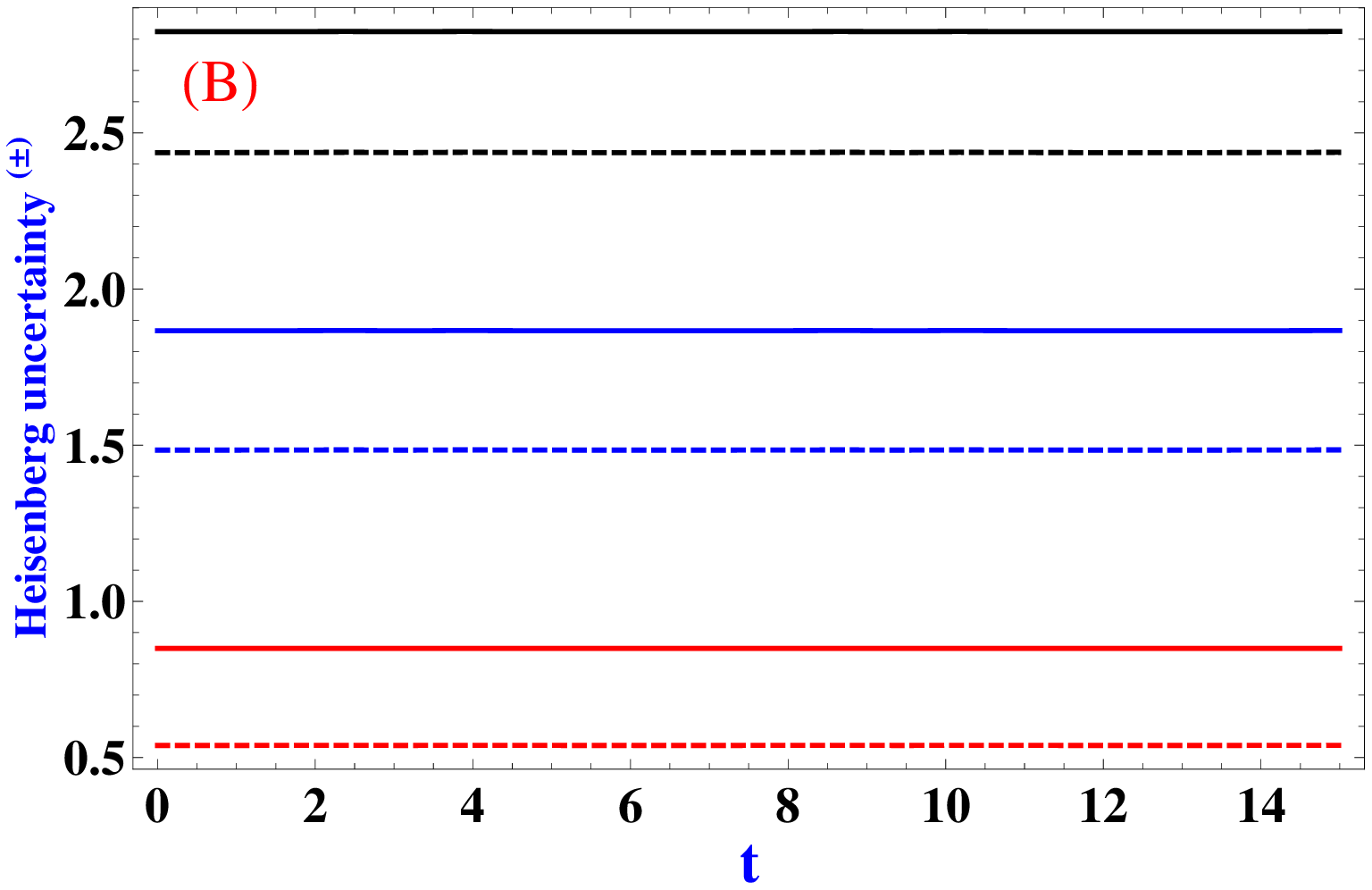}
\caption{\label{Fig.Compare.uncer.xp} Comparison of the uncertainty product for (A) 
$L(t)=\pi(2+\sin t)$, (B) $L(t)=\f{A_1\pi}{\sqrt{1+B_1\cos\om t}}$. The solid curves represent the $ (-)$ sector while the dashed curves represent the $(+)$ sector with red $(n=0)$, blue $(n=1)$ and black $(n=2)$. The parameter values are $A=5, B=3.4, A_1=1; B_1=0.5, \omega=1$.}
\end{figure}

Finally we consider the uncertainty product and  present plots of the uncertainty product  at different times in Fig \ref{Fig.Compare.uncer.xp}. From Fig \ref{Fig.Compare.uncer.xp}  we see that the uncertainty relation (\ref{uncer}) always holds good for all values of $t$. Secondly, it may be observed that the uncertainty product of the $(+)$ sector is always lower than that of the $(-)$ sector for all values of $n$ considered here and this may again be attributed to the nature of the solutions of the $(+)$ sector. These observations are true for both profiles of the moving boundaries.

\section{Conclusion}
In this paper we have constructed the time dependent versions of the P\"oschl-Teller potential and its supersymmetric partner, namely, the rationally extended P\"oschl-Teller potential and obtained exact solutions using separation of variable technique. Using these solutions, one of which are expressed in terms of $X_1$ Jacobi exceptional orthogonal polynomials, we have computed several quantities, for example, the average energy, expectation values etc. and examined their similarity/differences. It may be noted that a host of potentials whose solutions are given in terms of exceptional orthogonal polynomials of different types have been constructed in recent years \cite{sasaki,sasaki1}. We feel it would be interesting to treat the potentials considered here as well as others of different types \cite{sasaki,sasaki1} using other methods like the method of invariants and examine the differences, if any, with the present approach. In addition other types of time dependent boundary conditions, for example, when both boundaries are moving, may also be used to study the above mentioned problems.\\

{\bf Conflict of interest }: The authors do not have any conflict of interest.\\

{\bf Author's contributions} : All authors contributed equally to this work.\\


\begin{thebibliography}{}
	\bibitem{TD.Schrodinger} J. R. Ray, Phys. Rev. A {\bf 26} (1982) 729-733.
	\bibitem{TD.Schrodinger.2} M. J. Englefield, J. Phys. A: Math. Gen. {\bf 20} (1987) 593-600.	
	\bibitem{TD.Schrodinger.3} I. H. Duru, J. Phys. A: Math. Gen. {\bf 22} (1989) 4827-4833.
	\bibitem{TD.Schrodinger.4} V. V. Donodov, V. I. Manko, D. E. nikonov, Phys. Lett. A {\bf 162} (1992) 359-364.	
	\bibitem{TD.Schrodinger.5} V. G. Bagrov, B. F. Samsonov, Phys. Lett. A {\bf 210} (1996) 60-64.
	\bibitem{TD.Schrodinger.6} F. Finkel, A. Gonzalez-L\'opez, N. Kamran, M. A. Rodrguez, J. Math. Phys. {\bf 40} (1999) 3268-3274.
	\bibitem{TD.Schrodinger.7} B. F. Samsonov, L. A. Shekoyan, Phys. Atomic Nuclei {\bf 63} (2000) 657-660.
	\bibitem{TD.Schrodinger.8} B. F. Samsonov, Proceedings Inst. Math. NAS Ukrain {\bf 43} (2002) 520-529.
	\bibitem{TD.Schrodinger.9} S. Carrasco, J. Rogan, J. A. Valdivia, Scientific Report {\bf 7} (2017) 13217.
	\bibitem{Fermi} E. Fermi, Phys. Rev. {\bf 75} (1949) 1169-1174.
\bibitem{MB.First} J. Crank, {\it Free and moving boundary problems}, Clarendon Press, Oxford (1984).
\bibitem{moving.wall} S. W. Doescher, M. H. Rice, Am. J. Phys. {\bf 37} (1969) 1246-1249.
\bibitem{moving.wall.2}A. Munier, J. R. Burgan, M. Feix, E. Fajalkow, J. Math. Phys. {\bf 22} (1981) 1219-1223.
\bibitem{moving.wall.3} D. N. Pinder, Am. J. Phys. {\bf 58} (1990) 54-58.
\bibitem{moving.wall.4} A. J. Makowski, S. T. Dembinski, Phys. Lett. A {\bf 154} (1991) 217-220.
\bibitem{moving.wall.5} A. J. Makowski, P. Peplowski, Phys. Lett. A {\bf 163} (1992) 142-151.
\bibitem{moving.wall.6} A. J. Makowski, J Phys A {\bf 25} (1992) 3419-3426.
\bibitem{moving.wall.7} V. V. Dodonov, A. B. Klimov, D. E. Nikinov, J Math. Phys. {\bf 34} (1993) 3391-3404.
\bibitem{moving.wall.8} J. M. Cerver\'o, J. D. Lejareta, Euro. Phys. Lett. {\bf 45} (1999) 6-12.
\bibitem{moving.wall.9} J. D. Lejarreta, J. Phys. A {\bf 32} (1999) 4749-4759.
\bibitem{moving.wall.10} D.J. Fern\'andez C., J. Negro, L.M. Nieto, Phys. Lett. A {\bf 275} (2000) 338.
\bibitem{moving.wall.11} L. Ling, L. Bo-Zang, Phys. Lett. A {\bf 291} (2001) 190-206.
\bibitem{moving.wall.12} C. Y\"uce, Phys. Lett. A {\bf 321} (2004) 291-294.
\bibitem{moving.wall.13} C. Y\"uce, Phys. Lett. A {\bf 327} (2004) 107-112.
\bibitem{moving.wall.14} T. K. Jana, P. Roy, Phys. Lett. A {\bf 372} (2008) 2368-2373.
\bibitem{moving.wall.15} M. L. Glasser, J. Mateo, J. Negro, L. M. Nieto, Chaos,
Solitons and Fractals {\bf 41} (2009) 2067-2074.
\bibitem{moving.wall.16} O. Foj\'on, M. Gadella, L. P. Lara, Comput. Math. Appl. {\bf 59} (2010) 964-976.
\bibitem{moving.wall.17} P. Patra, A. Dutta, J. P. Saha, Pramana J. Phys. {\bf 80} (2013) 21-30.
\bibitem{moving.wall.18} A. Contreras-Astorga, V. Hussin, Integrability, Supersymmetry and Coherent States, pg 2885-299. CRM Series in Mathematical Physics. Springer, Cham, S. Kuru., J. Negro, L. Nieto (eds); arXiv:1901.04606v1 [quant-ph] 14 Jan 2019.
\bibitem{invariant} H. R. Lewis, Jr., W. B. Riesenfeld, J. Math. Phys. {\bf 10} (1969) 1458-
\bibitem{symmetry.separation} W. Miller, Jr., {\it Symmetry and separation of variables} London, Addison (1977).
\bibitem{symmetry.separation.2} W. Miller, Jr., E. G. Kalnins, J. Math. Phys. {\bf 28} (1987) 1005-1015.
\bibitem{symmetry.separation.3} J. Rogers, D. Spector, Phys. Lett. A {\bf 170} (1992) 344-346.
\bibitem{symmetry.separation.4} C.J. Efthimiou, D. Spector, Phys. Rev. A {\bf 49} (1994) 2301-2311.
\bibitem{eop} D. G\'omez-Ullate, N. Kamran and R. Milson, J. Math. Analysis and App. {\bf 359} (2009) 352.
\bibitem{eop1} D. G\'omez-Ullate, N. Kamran and R. Milson, J. Phys. {\bf A43} (2010) 434016.
\bibitem{eseop} C. Quesne, J. Phys. {\bf A41} (2008) 392001.
\bibitem{dutta} D. Dutta and P. Roy, J. Math. Phys. {\bf 51} (2010) 042101.
\bibitem{axel} A. Schulze-Halberg and B. Roy, J. Math. Phys. {\bf 55} (2014) 123506.
\bibitem{quesne} C. Quesne, SIGMA {\bf 5} (2009), 084.
\bibitem{nieto} D.J. Fern\'andez C., J. Negro, L.M. Nieto, Phys. Lett. A {\bf 275} (2000) 338.
\bibitem{susy1} F. Cooper, A. Khare, U. Sukhatme, Supersymmetry in Quantum Mechanics, World Scientific, 2001.
\bibitem{susy2} G. Junker, Supersymmetric Methods in Quantum and Statistical Physics, Springer, New York, 1996.
\bibitem{RMS} M. J. W. Hall, Phys. Rev. A {\bf 59} (1999) 2602-2615.
\bibitem{RMS.2} P. S\'anchez-Moreno, J. S. Dehesa, D. Manzano, R. J. Y\'a\~nez, J. Comput. Appl. Math. {\bf 233} (2010) 2136-2148.
\bibitem{average.energy} C. Yuce, Phys. Lett. A {\bf 336} (2005) 290-294.
\bibitem{sasaki} S. Odake and R. Sasaki Phys. Lett. {\bf B702} (2011) 164.
\bibitem{sasaki1} R. Sasaki, S. Tsujimoto and A. Zhedanov, J. Phys. {\bf A43} (2010) 315204.
\end{thebibliography}
\end{document}